\documentclass[aps,prl,notitlepage,twocolumn,superscriptaddress,longbibliography,nofootinbib]{revtex4-1}
\pdfoutput=1
\usepackage[utf8]{inputenc}
\usepackage{amsmath,amsthm,amssymb,amsfonts}
\usepackage{scalerel}
\usepackage{mathtools}
\usepackage{graphicx}
\usepackage{subfigure}
\usepackage[normalem]{ulem}
\usepackage[colorlinks = true,
            linkcolor = blue,
            urlcolor  = blue,
            citecolor = blue,
            anchorcolor = blue]{hyperref}
\usepackage{mathrsfs}
\usepackage{bbold}
\usepackage{units}
\usepackage{bm}
\usepackage{braket}
\usepackage{makecell}
\usepackage{enumitem}
\usepackage{upgreek}
\usepackage{blindtext}
\usepackage{verbatim}
\usepackage{algorithm}
\usepackage[noend]{algpseudocode}

\usepackage[dvipsnames]{xcolor}
\usepackage{bbm}
\usepackage[bb=boondox]{mathalfa}
\usepackage{array}
\usepackage{multirow}
\usepackage{tabularx}
\usepackage{float}
\usepackage{dcolumn}
\usepackage{slashed}
\usepackage{braket}
\usepackage{verbatim}
\usepackage{multirow}
\usepackage{resizegather}
\usepackage{titlesec}
\usepackage[toc,page]{appendix}
\usepackage[export]{adjustbox}

\begin{document}

\title{{Probing quantum chaos through singular-value correlations \\in sparse non-Hermitian SYK model}}

\author{Pratik Nandy\,\,\href{https://orcid.org/0000-0001-5383-2458}
{\includegraphics[scale=0.05]{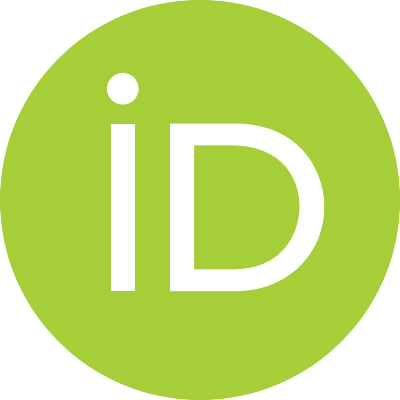}}}
\email{pratik@yukawa.kyoto-u.ac.jp}
\affiliation{Center for Gravitational Physics and Quantum Information, Yukawa Institute for Theoretical Physics,\\ Kyoto University, Kitashirakawa Oiwakecho, Sakyo-ku, Kyoto 606-8502, Japan}
\affiliation{RIKEN Interdisciplinary Theoretical and Mathematical Sciences Program (iTHEMS),
Wako, Saitama 351-0198, Japan}
\author{Tanay Pathak\,\,\href{https://orcid.org/0000-0003-0419-2583}
{\includegraphics[scale=0.05]{orcidid.pdf}}}
\email{pathak.tanay@yukawa.kyoto-u.ac.jp}
\affiliation{Center for Gravitational Physics and Quantum Information, Yukawa Institute for Theoretical Physics,\\ Kyoto University, Kitashirakawa Oiwakecho, Sakyo-ku, Kyoto 606-8502, Japan}
\author{Masaki Tezuka\,\,\href{https://orcid.org/0000-0001-7877-0839}
{\includegraphics[scale=0.05]{orcidid.pdf}}\,}
\email{tezuka@scphys.kyoto-u.ac.jp}
\affiliation{Department of Physics, Kyoto University, Kitashirakawa Oiwakecho, Sakyo-ku, Kyoto 606-8502, Japan}

\begin{abstract}

Exploring the spectral properties of non-Hermitian systems presents a substantial theoretical challenge due to the presence of a complex eigenvalue spectrum. Singular values for such systems are inherently real and non-negative, and the techniques used for Hermitian systems can be used with ease.
As a prototypical example of such systems, we investigate the singular-value spectrum of a non-Hermitian extension of the sparse Sachdev-Ye-Kitaev (SYK) model, a solvable toy model of quantum chaos and quantum gravity with significant interest in digital quantum simulation.
Our findings reveal a congruence between the statistics of singular values and those of the analogous Hermitian Gaussian ensembles. An increase in sparsity results in the model deviating from its chaotic behavior, a phenomenon precisely captured by both short and long-range correlations in the singular-value spectrum. These findings indicate the existence of a critical sparsity threshold, beyond which the chaotic nature of the model breaks down, suggesting a potential collapse of holographic duality in such systems.
\end{abstract}

~~~~~~~~~~~~YITP-24-70, RIKEN-iTHEMS-Report-24

\maketitle

\emph{Introduction}: Spectral statistics serve as a critical tool for probing the energy levels in quantum systems, offering insights into the dynamics within quantum chaos and Random Matrix Theory (RMT) \cite{mehta1991random, BGS, Wigner1, Wigner2, Dyson1962a, Dyson1962b, Berry_Tabor, Berry_Tabor_conj}. Short-range level statistics such as the ratio of consecutive level spacings \cite{Atas2013distribution} and the long-range correlations such as the Spectral Form Factor (SFF) \cite{PhysRevLett.75.4389, Guhr:1997ve,
Leviandier1986, WilkieBrumer1991, Brezin1997, Cotler2017} reveal the symmetries and randomness in the system. The latter has also been understood from the semiclassical gravity \cite{Saad:2018bqo} and measured in many-body scenarios \cite{das2024proposal, Dong:2024yaf}. Analogously, the upsurge in research has been directed towards unraveling the quantum chaotic properties of open quantum systems \cite{Denisov:2018nif, Yang:2024boo, Nandy:2024htc} described by non-Hermitian Hamiltonians \cite{Bender:1998ke, Katsurasissipative, Ashida:2020dkc, Cornelius:2021ccu, Matsoukas-Roubeas:2022odk}. This connotes a departure from conventional methods suited for Hermitian systems and necessitates new methods for spectral analysis \cite{ComplexspacingProsen, ProsenDSFF, Li:2024uzg, Cipolloni:2023mwh, GarcianonHSYK, Xu:2020wky, Zhou:2023qmk}. One such method approaches through the Singular Value Decomposition (SVD) \cite{KawabataSVD23}, which facilitates a straightforward generalization of the frameworks used for Hermitian systems with real and non-negative singular values. Consequently, this enables us to define the singular spacing ratios \cite{KawabataSVD23} and singular form factor ($\upsigma$FF) \cite{ChenuSVD} in a much more elegant manner capturing transitions from integrability to chaos, many-body localization \cite{HamazakiKawabataUeda, ChenuSVD}, and the non-Hermitian skin effect \cite{nonHskin1, nonHskin2, Hamanaka:2024njv}. Outside this domain, SVD has also found applications in generalizing entanglement entropy for non-Hermitian transition matrices in AdS/CFT correspondence \cite{Parzygnat:2023avh}.

In this paper, we utilize the SVD framework to study the quantum chaotic properties and their transitions in the sparse, non-Hermitian version of the Sachdev-Ye-Kitaev (SYK) model \cite{PhysRevLett.70.3339, Kittu}. Standing as a quintessential archetype for quantum chaos and possessing a dual gravitational theory at low temperatures \cite{Maldacena:2016hyu}, the SYK model (see See \cite{RevModPhys.94.035004, Rosenhaus:2018dtp} for comprehensive reviews) can be adapted into sparse variants \cite{Xu:2020shn, Tezuka:2022mrr,  Anegawa:2023vxq, Caceres:2021nsa, Caceres:2023yoj}, maintaining its chaotic nature up to a critical sparsity threshold \cite{Garcia-Garcia:2020cdo, Orman:2024mpw}. This adaptation holds theoretical considerations in holography as well as the practical significance for the deployment of the SYK model on quantum processors \cite{Jafferis2022, Kobrin:2023rzr}, where managing a sparse matrix is more viable. Our focus is particularly drawn to the influence that sparsity exerts on the average singular value spacing ratio, denoted by $\langle r_\upsigma \rangle$, and the $\upsigma$FF, which serves as an extension of the conventional $\braket{r}$-value \cite{Atas2013distribution} and the SFF in Hermitian systems, respectively. Additionally, we introduce \emph{singular complexity}, inspired by spectral complexity, a boundary dual to the volume of the Einstein-Rosen bridge in gravitational theories \cite{Iliesiu:2021ari}. Our findings suggest that these metrics proficiently encapsulate the transition from chaos to integrability in the non-Hermitian SYK model, as we outline in the following sections.

\emph{Non-Hermitian sparse SYK model:} 
The prototypical model we consider is the $4$-body non-Hermitian sparse SYK model (nSYK) \cite{PhysRevLett.70.3339, Kittu}, given by the following Hamiltonian
\begin{align}
  H_{\mathrm{nSYK}}^{\mathrm{sparse}} =   \sum_{1 \leq a < b < c < d \leq N} x_{abcd} (J_{abcd} + i M_{abcd}) \,\psi_{a} \psi_{b} \psi_{c} \psi_{d}\,. \label{nhsykhsparse}
\end{align}
where $J_{abcd}$ and $M_{abcd}$ are the independent random couplings drawn from the Gaussian ensemble with zero mean and variance $\braket{J_{abcd}^2} = \braket{M_{abcd}^2} = 6/(p N^3)$ (see \cite{Lau:2020qnl, Ozaki:2024wpj} for clean SYK model), and $p$ is the sparsity parameter. The couplings $M_{abcd}$ explicitly break the Hermiticity of the Hamiltonian \cite{Xu:2020shn, Garcia-Garcia:2020cdo}. The variables $\psi_{k}$ are the Majorana fermions obeying Clifford algebra $\{\psi_a, \psi_b\} = \delta_{ab}$. Additionally, the random variables $x_{ijkl} \in \{0,1\}$ introduce an element of sparsity $p$ to the Hamiltonian.  The parameter $p$ governs the probability of $x_{abcd}$ being unity, thus determining the sparsity level within the Hamiltonian. A fully \emph{dense} model is characterized by  $x_{abcd}=1$ or $p=1$ for all combinations of $\{a,b,c,d\}$, aligning with the original nSYK Hamiltonian \cite{Garcia-Garcia:2020ttf, Garcia-Garcia:2021rle, Garcia-Garcia:2022xsh, Cipolloni:2022fej}. Conversely, a completely sparse model would have $x_{abcd} = 0$ or $p = 0$ for all combinations, representing the other extreme \cite{Xu:2020shn}. The addition of sparsity effectively reduces the total number of independent terms in the Hamiltonian from $O(N^4)$ to $O(p N^4) \sim O(kN)$, where $k$ is defined in \eqref{pfit}. Yet, the probabilistic nature of $x_{abcd}$ implies that the exact number of non-zero terms in $H_{\mathrm{nSYK}}^{\mathrm{sparse}}$ is not fixed but rather determined by the probability $p$. We follow \cite{Orman:2024mpw} without fixing the total number of terms in the sparse Hamiltonian in each realization.

\emph{Singular value decomposition and singular value statistics:} In Hermitian systems, the eigenspectrum of the Hamiltonian is a key indicator of chaotic dynamics. The eigenvalues $E_n$ are real, allowing for the definition of consecutive level spacings $s_n = E_{n+1} - E_n$, as an ordered list. For large Hilbert spaces, the average consecutive level spacing ratio $r_n$ is defined as $r_n = \mathrm{min}(s_{n}, s_{n+1})/\mathrm{max}(s_{n}, s_{n+1})$ with $\braket{r} = \mathrm{mean} (r_n)$ being the \emph{average $r$ value} \cite{Atas2013distribution}. Its usefulness lies in simplifying the numerical analysis by bypassing the need to unfold the spectrum for level statistics. While $\braket{r}$-value is well-defined for Hermitian systems, it falls short in applying for non-Hermitian systems where eigenvalues are complex, and thus an ordered sequence of level spacings is not feasible. In such scenarios, one may consider complex spacing ratios \cite{ComplexspacingProsen} and correspondingly the statistics of complex eigenvalues. However, an alternate proposal suggests focusing on singular values derived from the singular value decomposition (SVD) of the Hamiltonian \cite{KawabataSVD23}. SVD is expressed as $H = U \Sigma V^{\dagger}$, where $U$ and $V$ are unitary matrices, and $\Sigma$ is a diagonal matrix containing the real and non-negative singular values $\sigma_i$ \cite{Nielsen_Chuang_2010}. In Hermitian matrices, singular values are simply the absolute values of the eigenvalues, whereas, in non-Hermitian matrices, their relationship is more complex \cite{allard2024correlation}.

The singular values of a non-Hermitian matrix $H$ can be computed by the \emph{Hermitization} process \cite{FEINBERG1997579}, which involves constructing a new matrix $\mathsf{H}$, twice the dimension of $H$. This larger matrix incorporates both $H$ and its Hermitian conjugate $H^{\dagger}$ in a block structure \cite{FEINBERG1997579}
\begin{align}
    \mathsf{H} = \begin{pmatrix}
\mathbb{0} & H \\
H^{\dagger} & \mathbb{0}
\end{pmatrix} ~~ \Rightarrow ~~\mathsf{H}^2 = \begin{pmatrix}
H H^{\dagger} & \mathbb{0} \\
\mathbb{0} & H^{\dagger}H
\end{pmatrix}\,.
\end{align}
The eigenvalues of $\mathsf{H}$ are $\{\pm \sigma_n\}$, precisely containing the singular values of $H$. This is equivalent to constructing an effective Hamiltonian  $H_{\mathrm{eff}} = \sqrt{ H^{\dagger} H}$ or $H_{\mathrm{eff}} = \sqrt{H H^{\dagger}}$ so that the eigenvalues of $H_{\mathrm{eff}}$ represent the singular values of $H$ \cite{FEINBERG1997579}. Such construction of a Hermitized matrix resembles the formulation of the Wishart SYK model and its supersymmetric version, comprised of non-Hermitian conserved charges or supercharges \cite{Li:2017hdt, Sa:2021rwg}.

Given the singular values $\{\sigma_i\}$ of the corresponding non-Hermitian Hamiltonian $H$, one analogously defines $\lambda_n = \sigma_{n+1} - \sigma_n$ as the \emph{singular value spacings}. Consequently, the \emph{singular-value-spacing ratio} $r_{\upsigma,n}$ is defined analogous to the $r$-ratio as \cite{KawabataSVD23}
\begin{align}
    r_{\upsigma,n} = \frac{\mathrm{min}(\lambda_{n}, \lambda_{n+1})}{\mathrm{max}(\lambda_{n}, \lambda_{n+1})}\,, ~~~~~ \braket{r_{\upsigma}} = \mathrm{mean} (r_{\upsigma,n})\,. \label{sigmav}
\end{align}
Referred to as the  $\braket{r_{\upsigma}}$-value, they adhere to statistical distributions akin to those observed for non-Hermitian Hamiltonians, similar to the $\braket{r}$-value distributions for Hermitian matrices.  This framework has been instrumental in addressing the $38$-fold symmetry classification in non-Hermitian systems \cite{KawabataSVD23}.

\begin{table}
\begin{tabular}{|l|l|l|l|l|l|l|l|l|}
\hline
\,$\mathrm{System}$   & $N = 20$ & $N = 22$ & $N = 24$ & $N = 26$ & $N = 28$ & $N = 30$\\ \hline
$\langle r \rangle_{\mathrm{RMT}}$ &  0.6744    & 0.5996   & 0.5307   & 0.5996   & 0.6744 & 0.5996   \\ \hline
$\langle r_\upsigma \rangle_{\mathrm{nSYK}}$ & 0.6744   & 0.5997   & 0.5307   & 0.5996   & 0.6745 & 0.5995   \\ \hline

\end{tabular}
\caption{The $\braket{r_\upsigma}$-values for the non-Hermitian \emph{dense} SYK model for different system sizes $N = 20 \,(10000)$, $N = 22 \,(6000)$, $N = 24 \,(2000)$, $N = 26 \,(1000)$, $N = 28\, (100)$, and $N = 30\,(50)$, where the parenthesis includes the number of samples taken. Based on the $\braket{r_\upsigma}$ values, we identify the periodicity as $N$ mod 8 = 0 (GOE), 2, 6 (GUE), and 4 (GSE) \cite{You:2016ldz}. These GOE, GUE, and GSE correspond to the eigenvalue statistics of the Hermitian matrices.} The values are in precise agreement (within numerical accuracy) with the $\braket{r}_{\mathrm{RMT}}$-values for large $N$ results in \cite{Atas2013distribution}. \label{Tab1}
\end{table}

\begin{figure}[t]
   \centering
\includegraphics[width=0.4\textwidth]{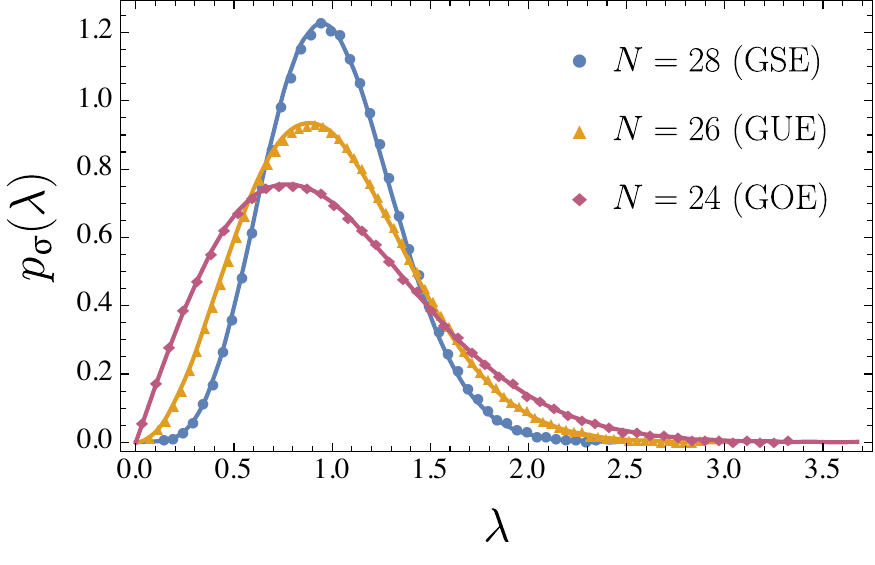}
\caption{The consecutive singular level-spacing distribution in the non-Hermitian dense SYK model for $N = 24\,(2000)$ (GOE), $26\,(1000)$ (GUE), and $28\,(100)$ (GSE) with the corresponding realizations shown in the brackets. The solid lines represent distribution for GOE, GUE, and GSE respectively taking matrices of size $10^4 \times 10^{4}$ with averaging over 1000 realizations.} \label{fig:SVDhistogramN24}
\end{figure}

Table\,\,\ref{Tab1} showcases the $\braket{r_\upsigma}$-value for the \emph{dense} non-Hermitian SYK model for varying system sizes $N$, representing the total number of fermions. The data is juxtaposed with corresponding eigenvalue statistics of Hermitian Gaussian ensembles, revealing a striking correlation emphasizing the utility of the $\braket{r_\upsigma}$-value in non-Hermitian systems. Similar to its Hermitian counterpart, we observe an $N$ mod 8 periodicity using singular value statistics: $N$ mod 8 = 0 corresponds to GOE, $N$ mod 8 = 2 and 6 corresponds to GUE, and $N$ mod 8 = 4 corresponds to GSE for the eigenvalue statistics of the Hermitian matrices. Figure \ref{fig:SVDhistogramN24} correspondingly shows the consecutive \emph{singular level-spacing distribution} $p_{\sigma}(\lambda)$ for $N = 24$ (GOE), $N = 26$ (GUE) and $N = 28$ (GSE) respectively, which are \emph{approximated} by the analogous Hermitian Wigner-Dyson distribution \cite{mehta1991random}.  Note that we are comparing the singular value statistics of non-Hermitian matrices to the eigenvalue statistics of Hermitian matrices. The singular value correlations in Hermitian matrices are generally weaker than the eigenvalue correlations and exhibit different spacing ratios \cite{KawabataSVD23}. However, the nSYK falls into the different non-Hermitian symmetry classes, which can also be used to show the validity of the singular value statistics \cite{Garcia-Garcia:2021rle}.

\begin{figure}[t]
   \centering
\includegraphics[width=0.85\linewidth]{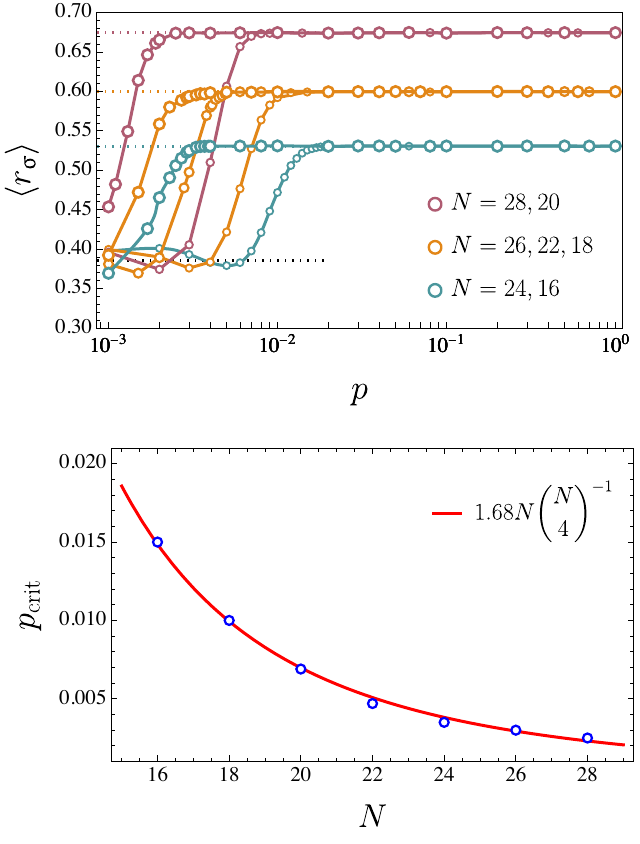}
\caption{(Top) Variation of $\braket{r_\upsigma}$-value with various level of sparsity $p$. The size of the circle varies with $N$ i.e., the largest $N$ corresponds to the leftmost plot for each color. At low sparsity, the $\braket{r_\upsigma}$ matches with the corresponding random matrix ensemble until a threshold $p_{\mathrm{crit}}$ is reached, after which $\braket{r_\upsigma}$ drops to the Poissonian value, indicating a hint of integrability. The estimated errors for the values obtained are smaller than the markers in the plot. (Bottom) The finite-size dependence of the threshold or the critical value $p_{\mathrm{crit}}$. It decreases as the system size increases. The decrease scales as $1/N^3$ with large system size, as given in \eqref{pfit}.} \label{fig:sigmavalsparse}
\end{figure}

\emph{SVD in sparse non-Hermitian SYK model:} We now turn our attention to the sparse variant of the model, which is governed by the sparsity parameter $p$. Figure \ref{fig:sigmavalsparse} illustrates the dependence of $p$ on $\braket{r_\upsigma}$-value for different system sizes. Each system size, representing a distinct ensemble, is depicted in a unique color. The dotted lines mark the expected values for Poisson statistics and the Hermitian ensemble, as specified in Table \ref{Tab1}. It is observed that the Hamiltonian \eqref{nhsykhsparse} retains its chaotic nature up to a critical level of sparsity denoted by $p_{\mathrm{crit}}$. This chaotic behavior is also corroborated by the $\braket{r_{\upsigma}}$-ratio presented in Table \ref{Tab1}. Consequently, we define a transition point $p_{\mathrm{crit}}$ beyond which the Hamiltonian \eqref{nhsykhsparse} ceases to exhibit chaotic dynamics. Such transition point is obtained by considering approximately $99\%$ of the corresponding $\braket{r_{\upsigma}}$ value at $p = 1$ \cite{Orman:2024mpw}. Notably, as the system size increases, this transition point migrates towards lower sparsity levels. When the sparsity is sufficiently low, the $\braket{r_{\upsigma}}$-value aligns with the Poisson limit, signifying the Hamiltonian's transition to complete integrability.Note that when a term of the Hamiltonian very weakly breaks a degeneracy, the resulting spectrum has $r_\sigma$ values close to zero, which can cause the $\braket{r_{\upsigma}}$ value to fall below the Poisson limit.

In conjunction with the above, the relationship between the critical sparsity $p_{\mathrm{crit}}$ and the system size $N$ is also noteworthy. The critical value diminishes as the system size increases. This reduction follows a power-law trend, best described by fitting the data to the curve \cite{Orman:2024mpw}
\begin{align}
    p_\mathrm{crit} =  k N \binom{N}{4}^{-1} \approx \frac{24 k}{N^3} \,, \label{pfit}
\end{align}
in the large $N$ limit. where the decrease is proportional to $1/N^3$ and the constant $k$ is approximately $1.68$, as determined by the fitting process. This scaling behavior is consistent with that observed for the sparse Hermitian SYK model, as reported in \cite{Garcia-Garcia:2020cdo, Orman:2024mpw}.

\emph{Spectral and Singular form factor: }
Contrary to the singular value statistics, the SFF probes the long-range eigenvalue correlations. It is defined as the Fourier transform of two-point eigenvalue correlation pairs \cite{
Leviandier1986,WilkieBrumer1991,Brezin1997}. For systems exhibiting quantum chaos, the SFF displays a characteristic dip-ramp-plateau pattern. The initial dip, evident in the early-time regime, is a non-universal trait influenced by the detailed nature of the energy spectrum \cite{PhysRevLett.75.4389}. This dip extends until a timescale known as the
\emph{Thouless time} ($t_{\mathrm{Th}}$) \cite{Gharibyan:2018jrp, Nosaka:2018iat} or \emph{ramp time} \cite{Cotler2017} beyond which the energy difference surpasses the inverse of $t_{\mathrm{Th}}$, referred to as the \emph{Thouless energy} \cite{PhysRevLett.39.1167}. Subsequently, a ramp emerges at $t_{\mathrm{Th}}$, indicative of eigenvalue repulsion, a phenomenon that mirrors the SFF behavior predicted by RMT \cite{Brezin1997, Cotler2017} and underscores the chaotic dynamics of the Hamiltonian. Moreover, this ramp hints at the presence of a holographic dual to the corresponding theory. Beyond the \emph{Heisenberg time} \cite{Erdos2015, PhysRevLett.78.2280}, which scales inversely to the mean level spacing, the SFF reaches a plateau that depends on the system size. In contrast, integrable systems typically bypass the ramp stage and saturate directly from the dip to the plateau regime. SFF in sparse SYK models has been thoroughly investigated in recent literature \cite{Garcia-Garcia:2020cdo, Tezuka:2022mrr, Anegawa:2023vxq, Orman:2024mpw}. Several other generalizations for SFF in non-Hermitian and open systems have also been proposed
\cite{GarcianonHSYK, Xu:2020wky, Zhou:2023qmk}.

Analogous to the SFF, the singular form factor ($\upsigma$FF) is defined as \cite{ChenuSVD}
\begin{align}
    \upsigma\mathrm{FF}(t) = \frac{1}{L^2}\braket{|Z_{\upsigma}(it)|^2} = \frac{1}{L^2} \bigg\langle \Big|\sum_n e^{- i \sigma_n t} \Big|^2 \bigg\rangle\,,   \label{sigff}
\end{align}
where $Z_{\upsigma}(i t) =  \sum_n  e^{- i \sigma_n t}$ is the infinite-temperature partition function of singular values, which can be further expressed as the overlap between the time-evolved right or left singular vectors governed by the effective Hamiltonian $\sqrt{H^{\dagger} H}$ or $\sqrt{H H^{\dagger}}$ respectively \cite{ChenuSVD}. Here the angle brackets indicate an ensemble average, which is particularly relevant in the context of a disordered system since $\upsigma$FF is not self-averaging \cite{PhysRevLett.78.2280}.

\begin{figure}[t]
\hspace*{-0.65 cm}
\includegraphics[width=0.5\textwidth]{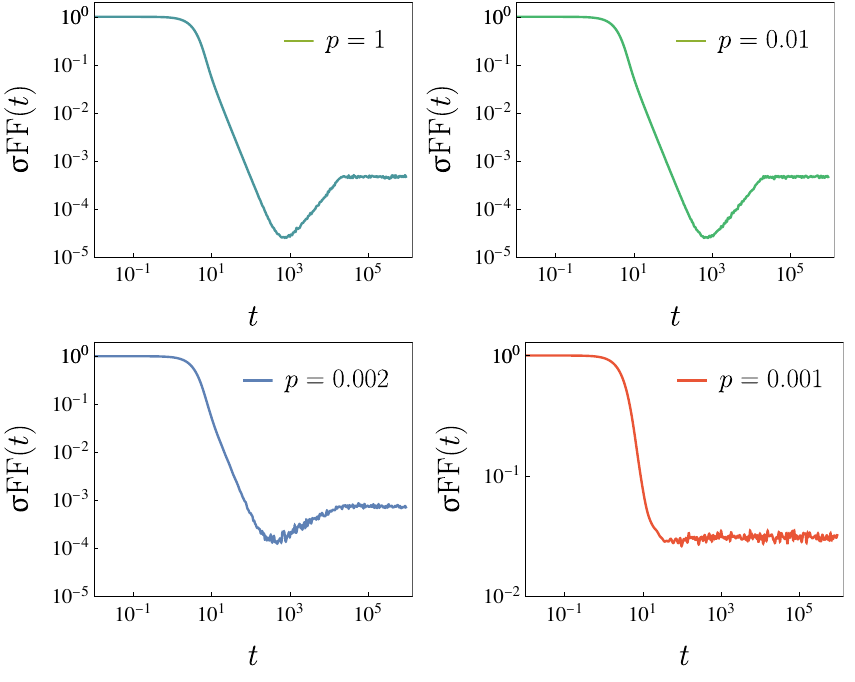}
\caption{Behavior of  $\upsigma$FF$(t)$ for various sparsity parameter $p$. The linear ramp ceases to exist as the sparsity increases. The system parameters are $N = 26$ (GUE) with $1000$ Hamiltonian realizations. Here we choose the Gaussian filter with $\alpha = 3.27$, which lies within the prescribed range in \cite{Orman:2024mpw}. 
} \label{fig:sigmaSFF}
\end{figure}

Figure \ref{fig:sigmaSFF} shows the distinct dip-ramp-plateau pattern exhibited by $\upsigma$FF for the Hamiltonian \eqref{nhsykhsparse}, across varying degrees of sparsity. In this work, we consider a system size of $N = 26$ and analyze the outcomes from $1000$ independent realizations of the Hamiltonian to ensure robust statistical analysis. To reduce the oscillations in the dip region, we use the filter function $Y_{\upsigma}(\alpha, t)$ \cite{Gharibyan:2018jrp} such that
\begin{align}
    \upsigma \mathrm{FF} (t) = \bigg\langle  \frac{|Y_{\upsigma}(\alpha, t)|^2}{|Y_{\upsigma}(\alpha, 0)|^2} \bigg\rangle \,,  \label{gaussfilter0}
\end{align}
Here the Gaussian filter function for the singular values is defined as $|Y_{\upsigma} (\alpha, t)|^2 = | \sum_n e^{-\alpha \sigma_n^2} \, e^{- i \sigma_n t}|^2$. Similar to the eigenspectrum
\cite{Gharibyan:2018jrp, Nosaka:2018iat}, it ensures that the dominant contribution is received from the bulk of the singular value spectrum. Here $\alpha$ is an $N$-dependent parameter that controls the Gaussianity of the spectrum. This filtering is performed for the usual spectrum, without unfolding it. For the latter case, an equivalent of \emph{connected unfolded} SFF \cite{Garcia-Garcia:2018ruf, Nosaka:2018iat} can be defined. We employ the filter function on the singular values by choosing $\alpha = 3.27$, which lies inside the range given in \cite{Orman:2024mpw} for $N =26$. Filter functions for SFF have also recently been shown to be associated with quantum channels \cite{Matsoukas-Roubeas:2023xge}.

A notable observation in Fig.\,\ref{fig:sigmaSFF} is a parallelism between the behavior of $\upsigma$FF in non-Hermitian systems and the SFF in Hermitian systems. This similarity underscores the efficacy of $\upsigma$FF as a robust indicator for quantum chaos in non-Hermitian frameworks.  In regimes of lower sparsity, the presence of the ramp is evident, signifying the retention of chaotic dynamics within this parameter space. It remains pronounced until a critical sparsity threshold is reached. Beyond this juncture, the linear ramp—a signature trait of non-integrable systems—ceases to manifest, indicating a shift towards integrability and a departure from chaotic behavior.

\begin{figure}
   \centering
\includegraphics[width=0.39\textwidth]{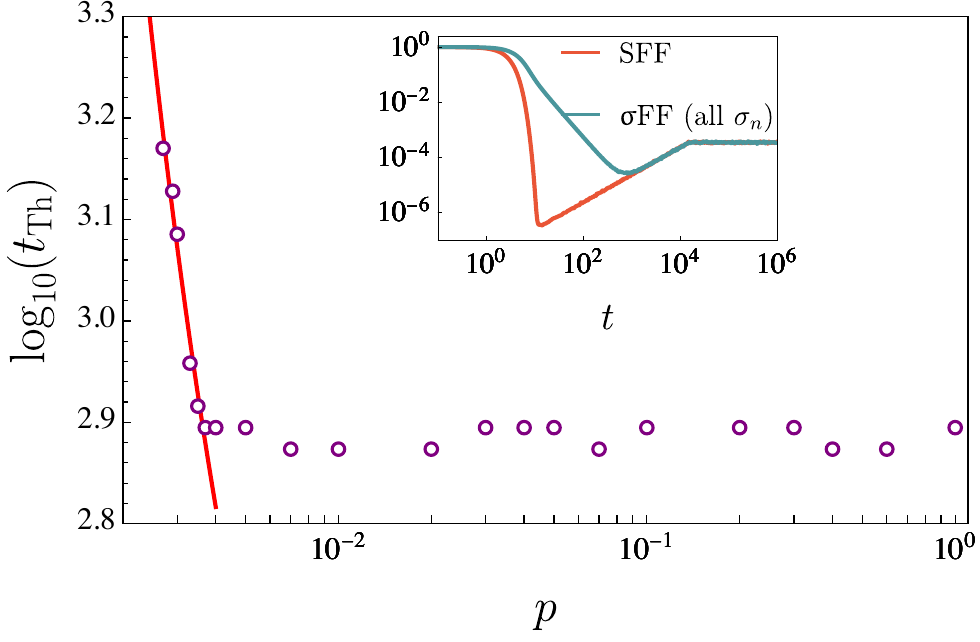}
\caption{(Main plot) Thouless time (ramp time) with respect to sparsity parameter. The red line is the fit $t_{\mathrm{Th}} \approx a/p^b + c$ with $a = 1.923$, $b = 0.8767$ and $c = 0.007$. (Inset) The behavior of SFF and the $\upsigma$FF for the Hermitian dense SYK$_4$ model. The system parameters are $N =26$ with $1000$ random Hamiltonian realizations taken. The $\upsigma$FF (green line) is computed considering all the positive singular values. It reduces to the SFF after taking the appropriate rescaling of the singular vectors and singular values, which otherwise equals the eigenvalues (with appropriate signs).} \label{fig:Thoulesstimevsp}
\end{figure}

Figure \ref{fig:Thoulesstimevsp} shows the variation of Thouless time ($t_{\mathrm{Th}}$), for $N= 26$, with sparseness parameter $p$. The $t_{\mathrm{Th}}$ is calculated by considering the time at which the $\upsigma \mathrm{FF}$ of the system considered first intersects the SFF of the GUE random matrices \cite{Cotler2017}. To determine the intersection point precisely, we consider the fractional error function $\epsilon(t) := \Big|\frac{\upsigma \mathrm{FF} (t) - \upsigma_{\mathrm{ramp}}(t)}{\upsigma_{\mathrm{ramp}} (t)}\Big|$
\cite{Gharibyan:2018jrp} indicating the deviation of the computed $\upsigma \mathrm{FF}$ to the linear-fitted function for the ramp $\upsigma_{\mathrm{ramp}} (t)$ for $p = 1$. We consider the value $t$ as the Thouless time or ramp time $t_{\mathrm{Th}}$, where the fractional error $\epsilon(t)$ reaches $20\%$. The red line
shows the fitting $t_{\mathrm{Th}} \approx a/p^b + c$ with $a = 1.923$, $b = 0.8767$ and $c = 0.007$. Curiously, this fitting follows closely to the $1/p$ scaling in contrast to the $1/p^2$ scaling in Hermitian systems \cite{Orman:2024mpw}. Such distinction is possible since the singular values and the eigenvalues are inherently different (see the Supplemental Material \cite{supp} S1). Yet, we do not rule out the possibility of the scaling of the exponent $b$ with $N$.

\emph{Spectral complexity for singular values - Singular complexity:} 
Within the framework of the AdS/CFT correspondence \cite{Maldacena:1997re, Witten:1998qj}, a particular spectral measure has been proposed to capture the essence of long-range spectral correlations. This measure, known as \emph{spectral complexity}, is proposed to be a dual quantity of the Einstein-Rosen bridge \cite{Iliesiu:2021ari} in the gravitational theory, serving as an analog for the computational and circuit complexity \cite{Jefferson:2017sdb}. In a similar spirit, we define \emph{singular complexity} at infinite temperature as 
\begin{align}
    C_{\upsigma}(t) = \frac{1}{L^2} \sum_{\sigma_i \neq \sigma_j} \left[\frac{\sin (t (\sigma_i - \sigma_j)/2)}{(\sigma_i - \sigma_j)/2}\right]^2\,,  \label{scTinf}
\end{align}
designed to be suitable for non-Hermitian systems. At early times, it grows quadratically, followed by a linear growth and plateau regime (see Supplemental Material \cite{supp} S2 for more details). The early time behavior mirrors the behavior of spread complexity \cite{Balasubramanian:2022tpr, Erdmenger:2023wjg} based on the Krylov space approach \cite{Parker:2018yvk}. The plateau value can be obtained analytically by taking the long-time average
\begin{align}
    \bar{C}_{\upsigma} = \lim_{t_{\mathrm{f}} \rightarrow \infty} \frac{1}{t_{\mathrm{f}}} \int_0^{t_{\mathrm{f}}} C_{\upsigma}(t) \, dt = \frac{2}{L^2} \sum_{\sigma_i \neq \sigma_j} \frac{1}{(\sigma_i - \sigma_j)^2}\,, \label{sigmasat}
\end{align}
which only depends on the difference between the singular values. In the Hermitian case, such saturation values have been studied for quantum billiards \cite{Camargo:2023eev} and mixed-field Ising model \cite{Camargo:2024deu}, capturing the \emph{spectral rigidity}. Integrable systems saturate at higher values compared to chaotic ones due to the presence of level repulsion,
as evident from the equivalent eigenspectrum of \eqref{sigmasat}.  

\begin{figure}
\hspace*{-0.9 cm}
\includegraphics[width=0.41\textwidth]{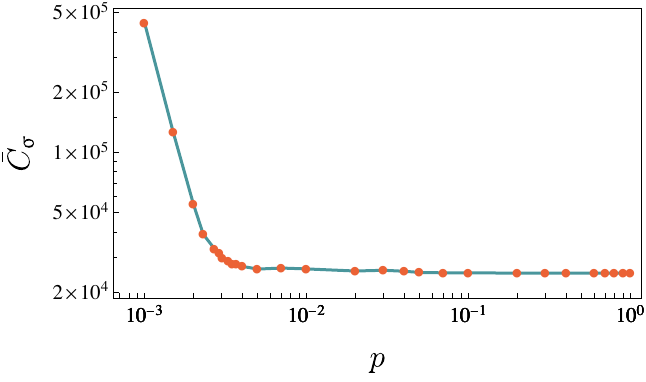}
\caption{Behavior of the saturation of the singular complexity with different sparseness for $N =26$ with $1000$ random Hamiltonian realizations.}\label{fig:SpecCN26}
\end{figure}

Figure \ref{fig:SpecCN26} depicts the saturation of singular complexity at varying levels of sparsity, which maintains a magnitude order of $O(10^4)$ until it encounters a critical threshold. Beyond this threshold, there is a marked escalation at lower sparsity levels signaling a level-repulsion between the singular values. This critical value is dependent on the system size $N$, and for $N = 26$, it occurs within the sparsity regime $0.003 \lesssim p_{\mathrm{crit}} \lesssim 0.004$. This value resembles the expected $\braket{r_\upsigma}$-value shown in Fig.\,\ref{fig:sigmavalsparse}.

Exhibiting chaotic dynamics is considered to be a crucial constraint that possesses a holographic dual geometry. While this discussion does not explore the specifics of such geometry, some proposals exist in the literature \cite{Garcia-Garcia:2020ttf, Garcia-Garcia:2022adg, Cai:2022onu, Cao:2024xaw}. The breakdown of chaotic nature suggests that such dual geometry corresponding to a non-Hermitian system may be compromised at low sparsity.

\emph{Conclusion and Outlook:} Drawing on the foundational work \cite{KawabataSVD23} on the applicability of SVD in non-Hermitian systems, our study addresses the quantum chaotic attributes of the sparse SYK model across varying levels of sparsity. The complex eigenvalues inherent to non-Hermitian Hamiltonians pose challenges to the generalization of traditional eigenvalue statistics. This complexity persists despite strides made with complex spacing ratios \cite{ComplexspacingProsen} and the DSFF \cite{ProsenDSFF, Li:2024uzg}. Nonetheless, recent advancements propose a novel perspective by focusing on singular values as opposed to eigenvalues within the non-Hermitian framework. Unlike their complex counterparts, singular values are inherently real and non-negative, aligning with the real eigenvalues in Hermitian systems upon appropriate alignment with singular vectors, as detailed in the Supplemental Material \cite{supp} S1.

Our investigation underscores that the statistical analysis of singular values, which elucidates short-range correlations, coupled with the $\upsigma$FF and singular complexity, illuminating long-range correlations among singular values, serve as potent indicators of the chaos within non-Hermitian systems. As such, exploring the holographic bulk dual of singular complexity may offer insights into a potential holographic counterpart for non-Hermitian systems, especially the effective range of $k$ in \eqref{pfit} in relation to $2d$ de Sitter (dS$_2$) gravity \cite{Garcia-Garcia:2022adg}, which we leave for future work. Moreover, these tools adeptly reflect the shift towards integrability as sparsity intensifies. Such transition differs from other chaotic-integrable transitions by two-point interactions observed in the SYK model \cite{Garcia-Garcia:2017bkg}. We envision that these features will be exhibited in other correlation measures such as higher spacing ratios \cite{Srivastava_2019, Santhanamhigherspacing18, Shir:2023olc} and the number variance \cite{GarcianonHSYK, Gharibyan:2018jrp, RevModPhys.80.1355, PhysRevD.94.126010}.

Parallel research avenues have explored the efficacy of sparse Hamiltonians \cite{Xu:2020shn, Tezuka:2022mrr} within quantum processors \cite{Jafferis2022, Kobrin:2023rzr}, maintaining chaotic dynamics to a substantial degree of sparsity. Notably, the SYK model retains its holographic duality even at minimal sparsity levels \cite{Garcia-Garcia:2020cdo, Orman:2024mpw}. Our study contributes to the burgeoning field of non-Hermitian systems in the realm of open-system dynamics, proposing a framework for the extrapolation of quantum chaotic properties within these systems. This might serve as the benchmark results for the simulation of sparse non-Hermitian systems on quantum processors.

\emph{Acknowledgements:} We would like to thank Hugo A. Camargo,  Adolfo del Campo, Aur\'elia Chenu, Antonio M. García-García, Luca V. Iliesiu, Yiyang Jia, Hosho Katsura, and Tadashi Takayanagi for fruitful discussions, comments, and suggestions on the draft. Numerical computations were performed in the \emph{Sushiki} workstation using the computational facilities of YITP. P.N. thanks the Berkeley Center for Theoretical Physics (BCTP), University of California for hosting him through the Adopting Sustainable Partnerships for Innovative Research Ecosystem (ASPIRE) program of Japan Science and Technology Agency (JST), Grant No. JPMJAP2318 during the final stages of the work. This work is supported by the Japan Society for the Promotion of Science (JSPS) Grants-in-Aid for Transformative Research Areas (A) ``Extreme Universe'' 
No.\,JP21H05190 (P.N.) and JP21H05182, JP21H05185 (M.T.). The Yukawa Research Fellowship of T.P. is supported by the Yukawa Memorial Foundation and JST CREST (Grant No. JPMJCR19T2). The work of M.T. was partially supported by the JSPS Grants-in-Aid for Scientific Research (KAKENHI) Grants No. JP20K03787 and JST CREST (Grant No. JPMJCR24I2).
\\
\newline
Authors' names are listed in alphabetical order.

\bibliography{references}

\begin{thebibliography}{94}%
\makeatletter
\providecommand \@ifxundefined [1]{%
 \@ifx{#1\undefined}
}%
\providecommand \@ifnum [1]{%
 \ifnum #1\expandafter \@firstoftwo
 \else \expandafter \@secondoftwo
 \fi
}%
\providecommand \@ifx [1]{%
 \ifx #1\expandafter \@firstoftwo
 \else \expandafter \@secondoftwo
 \fi
}%
\providecommand \natexlab [1]{#1}%
\providecommand \enquote  [1]{``#1''}%
\providecommand \bibnamefont  [1]{#1}%
\providecommand \bibfnamefont [1]{#1}%
\providecommand \citenamefont [1]{#1}%
\providecommand \href@noop [0]{\@secondoftwo}%
\providecommand \href [0]{\begingroup \@sanitize@url \@href}%
\providecommand \@href[1]{\@@startlink{#1}\@@href}%
\providecommand \@@href[1]{\endgroup#1\@@endlink}%
\providecommand \@sanitize@url [0]{\catcode `\\12\catcode `\$12\catcode
  `\&12\catcode `\#12\catcode `\^12\catcode `\_12\catcode `\%12\relax}%
\providecommand \@@startlink[1]{}%
\providecommand \@@endlink[0]{}%
\providecommand \url  [0]{\begingroup\@sanitize@url \@url }%
\providecommand \@url [1]{\endgroup\@href {#1}{\urlprefix }}%
\providecommand \urlprefix  [0]{URL }%
\providecommand \Eprint [0]{\href }%
\providecommand \doibase [0]{http://dx.doi.org/}%
\providecommand \selectlanguage [0]{\@gobble}%
\providecommand \bibinfo  [0]{\@secondoftwo}%
\providecommand \bibfield  [0]{\@secondoftwo}%
\providecommand \translation [1]{[#1]}%
\providecommand \BibitemOpen [0]{}%
\providecommand \bibitemStop [0]{}%
\providecommand \bibitemNoStop [0]{.\EOS\space}%
\providecommand \EOS [0]{\spacefactor3000\relax}%
\providecommand \BibitemShut  [1]{\csname bibitem#1\endcsname}%
\let\auto@bib@innerbib\@empty
\bibitem [{\citenamefont {Mehta}(1991)}]{mehta1991random}%
  \BibitemOpen
  \bibfield  {author} {\bibinfo {author} {\bibfnamefont {M.L.}\ \bibnamefont
  {Mehta}},\ }\href {https://books.google.co.jp/books?id=-sloQgAACAAJ} {\emph
  {\bibinfo {title} {Random Matrices}}}\ (\bibinfo  {publisher} {Academic
  Press},\ \bibinfo {year} {1991})\BibitemShut {NoStop}%
\bibitem [{\citenamefont {Bohigas}\ \emph {et~al.}(1984)\citenamefont
  {Bohigas}, \citenamefont {Giannoni},\ and\ \citenamefont {Schmit}}]{BGS}%
  \BibitemOpen
  \bibfield  {author} {\bibinfo {author} {\bibfnamefont {O.}~\bibnamefont
  {Bohigas}}, \bibinfo {author} {\bibfnamefont {M.~J.}\ \bibnamefont
  {Giannoni}}, \ and\ \bibinfo {author} {\bibfnamefont {C.}~\bibnamefont
  {Schmit}},\ }\bibfield  {title} {\enquote {\bibinfo {title} {Characterization
  of chaotic quantum spectra and universality of level fluctuation laws},}\
  }\href {\doibase 10.1103/PhysRevLett.52.1} {\bibfield  {journal} {\bibinfo
  {journal} {Phys. Rev. Lett.}\ }\textbf {\bibinfo {volume} {52}},\ \bibinfo
  {pages} {1--4} (\bibinfo {year} {1984})}\BibitemShut {NoStop}%
\bibitem [{\citenamefont {Wigner}(1955)}]{Wigner1}%
  \BibitemOpen
  \bibfield  {author} {\bibinfo {author} {\bibfnamefont {Eugene~P.}\
  \bibnamefont {Wigner}},\ }\bibfield  {title} {\enquote {\bibinfo {title}
  {{Characteristic Vectors of Bordered Matrices With Infinite Dimensions}},}\
  }\href {http://www.jstor.org/stable/1970079} {\bibfield  {journal} {\bibinfo
  {journal} {Annals of Mathematics}\ }\textbf {\bibinfo {volume} {62}},\
  \bibinfo {pages} {548--564} (\bibinfo {year} {1955})}\BibitemShut {NoStop}%
\bibitem [{\citenamefont {Wigner}(1957)}]{Wigner2}%
  \BibitemOpen
  \bibfield  {author} {\bibinfo {author} {\bibfnamefont {Eugene~P.}\
  \bibnamefont {Wigner}},\ }\bibfield  {title} {\enquote {\bibinfo {title}
  {{Characteristics Vectors of Bordered Matrices with Infinite Dimensions
  II}},}\ }\href {http://www.jstor.org/stable/1969956} {\bibfield  {journal}
  {\bibinfo  {journal} {Annals of Mathematics}\ }\textbf {\bibinfo {volume}
  {65}},\ \bibinfo {pages} {203--207} (\bibinfo {year} {1957})}\BibitemShut
  {NoStop}%
\bibitem [{\citenamefont {Dyson}(1962{\natexlab{a}})}]{Dyson1962a}%
  \BibitemOpen
  \bibfield  {author} {\bibinfo {author} {\bibfnamefont {Freeman~J.}\
  \bibnamefont {Dyson}},\ }\bibfield  {title} {\enquote {\bibinfo {title}
  {{Statistical Theory of the Energy Levels of Complex Systems. I}},}\ }\href
  {\doibase 10.1063/1.1703773} {\bibfield  {journal} {\bibinfo  {journal}
  {Journal of Mathematical Physics}\ }\textbf {\bibinfo {volume} {3}},\
  \bibinfo {pages} {140--156} (\bibinfo {year}
  {1962}{\natexlab{a}})}\BibitemShut {NoStop}%
\bibitem [{\citenamefont {Dyson}(1962{\natexlab{b}})}]{Dyson1962b}%
  \BibitemOpen
  \bibfield  {author} {\bibinfo {author} {\bibfnamefont {Freeman~J.}\
  \bibnamefont {Dyson}},\ }\bibfield  {title} {\enquote {\bibinfo {title}
  {{Statistical Theory of the Energy Levels of Complex Systems. II}},}\ }\href
  {\doibase 10.1063/1.1703774} {\bibfield  {journal} {\bibinfo  {journal}
  {Journal of Mathematical Physics}\ }\textbf {\bibinfo {volume} {3}},\
  \bibinfo {pages} {157--165} (\bibinfo {year}
  {1962}{\natexlab{b}})}\BibitemShut {NoStop}%
\bibitem [{\citenamefont {Berry}\ \emph {et~al.}(1977)\citenamefont {Berry},
  \citenamefont {Tabor},\ and\ \citenamefont {Ziman}}]{Berry_Tabor}%
  \BibitemOpen
  \bibfield  {author} {\bibinfo {author} {\bibfnamefont {Michael~Victor}\
  \bibnamefont {Berry}}, \bibinfo {author} {\bibfnamefont {M.}~\bibnamefont
  {Tabor}}, \ and\ \bibinfo {author} {\bibfnamefont {John~Michael}\
  \bibnamefont {Ziman}},\ }\bibfield  {title} {\enquote {\bibinfo {title}
  {Level clustering in the regular spectrum},}\ }\href {\doibase
  10.1098/rspa.1977.0140} {\bibfield  {journal} {\bibinfo  {journal}
  {Proceedings of the Royal Society of London. A. Mathematical and Physical
  Sciences}\ }\textbf {\bibinfo {volume} {356}},\ \bibinfo {pages} {375--394}
  (\bibinfo {year} {1977})}\BibitemShut {NoStop}%
\bibitem [{\citenamefont {Berry}\ and\ \citenamefont
  {Tabor}(1976)}]{Berry_Tabor_conj}%
  \BibitemOpen
  \bibfield  {author} {\bibinfo {author} {\bibfnamefont {M.~V.}\ \bibnamefont
  {Berry}}\ and\ \bibinfo {author} {\bibfnamefont {M.}~\bibnamefont {Tabor}},\
  }\bibfield  {title} {\enquote {\bibinfo {title} {Closed orbits and the
  regular bound spectrum},}\ }\href {http://www.jstor.org/stable/79042}
  {\bibfield  {journal} {\bibinfo  {journal} {Proceedings of the Royal Society
  of London. Series A, Mathematical and Physical Sciences}\ }\textbf {\bibinfo
  {volume} {349}},\ \bibinfo {pages} {101--123} (\bibinfo {year}
  {1976})}\BibitemShut {NoStop}%
\bibitem [{\citenamefont {Atas}\ \emph {et~al.}(2013)\citenamefont {Atas},
  \citenamefont {Bogomolny}, \citenamefont {Giraud},\ and\ \citenamefont
  {Roux}}]{Atas2013distribution}%
  \BibitemOpen
  \bibfield  {author} {\bibinfo {author} {\bibfnamefont {Y.~Y.}\ \bibnamefont
  {Atas}}, \bibinfo {author} {\bibfnamefont {E.}~\bibnamefont {Bogomolny}},
  \bibinfo {author} {\bibfnamefont {O.}~\bibnamefont {Giraud}}, \ and\ \bibinfo
  {author} {\bibfnamefont {G.}~\bibnamefont {Roux}},\ }\bibfield  {title}
  {\enquote {\bibinfo {title} {Distribution of the ratio of consecutive level
  spacings in random matrix ensembles},}\ }\href {\doibase
  10.1103/PhysRevLett.110.084101} {\bibfield  {journal} {\bibinfo  {journal}
  {Phys. Rev. Lett.}\ }\textbf {\bibinfo {volume} {110}},\ \bibinfo {pages}
  {084101} (\bibinfo {year} {2013})}\BibitemShut {NoStop}%
\bibitem [{\citenamefont {Agam}\ \emph {et~al.}(1995)\citenamefont {Agam},
  \citenamefont {Altshuler},\ and\ \citenamefont
  {Andreev}}]{PhysRevLett.75.4389}%
  \BibitemOpen
  \bibfield  {author} {\bibinfo {author} {\bibfnamefont {Oded}\ \bibnamefont
  {Agam}}, \bibinfo {author} {\bibfnamefont {Boris~L.}\ \bibnamefont
  {Altshuler}}, \ and\ \bibinfo {author} {\bibfnamefont {Anton~V.}\
  \bibnamefont {Andreev}},\ }\bibfield  {title} {\enquote {\bibinfo {title}
  {Spectral statistics: From disordered to chaotic systems},}\ }\href {\doibase
  10.1103/PhysRevLett.75.4389} {\bibfield  {journal} {\bibinfo  {journal}
  {Phys. Rev. Lett.}\ }\textbf {\bibinfo {volume} {75}},\ \bibinfo {pages}
  {4389--4392} (\bibinfo {year} {1995})}\BibitemShut {NoStop}%
\bibitem [{\citenamefont {Guhr}\ \emph {et~al.}(1998)\citenamefont {Guhr},
  \citenamefont {Muller-Groeling},\ and\ \citenamefont
  {Weidenmuller}}]{Guhr:1997ve}%
  \BibitemOpen
  \bibfield  {author} {\bibinfo {author} {\bibfnamefont {Thomas}\ \bibnamefont
  {Guhr}}, \bibinfo {author} {\bibfnamefont {Axel}\ \bibnamefont
  {Muller-Groeling}}, \ and\ \bibinfo {author} {\bibfnamefont {Hans~A.}\
  \bibnamefont {Weidenmuller}},\ }\bibfield  {title} {\enquote {\bibinfo
  {title} {{Random matrix theories in quantum physics: Common concepts}},}\
  }\href {\doibase 10.1016/S0370-1573(97)00088-4} {\bibfield  {journal}
  {\bibinfo  {journal} {Phys. Rept.}\ }\textbf {\bibinfo {volume} {299}},\
  \bibinfo {pages} {189--425} (\bibinfo {year} {1998})}\BibitemShut {NoStop}%
\bibitem [{\citenamefont {Leviandier}\ \emph {et~al.}(1986)\citenamefont
  {Leviandier}, \citenamefont {Lombardi}, \citenamefont {Jost},\ and\
  \citenamefont {Pique}}]{Leviandier1986}%
  \BibitemOpen
  \bibfield  {author} {\bibinfo {author} {\bibfnamefont {Luc}\ \bibnamefont
  {Leviandier}}, \bibinfo {author} {\bibfnamefont {Maurice}\ \bibnamefont
  {Lombardi}}, \bibinfo {author} {\bibfnamefont {R\'emi}\ \bibnamefont {Jost}},
  \ and\ \bibinfo {author} {\bibfnamefont {Jean~Paul}\ \bibnamefont {Pique}},\
  }\bibfield  {title} {\enquote {\bibinfo {title} {Fourier transform: A tool to
  measure statistical level properties in very complex spectra},}\ }\href
  {\doibase 10.1103/PhysRevLett.56.2449} {\bibfield  {journal} {\bibinfo
  {journal} {Phys. Rev. Lett.}\ }\textbf {\bibinfo {volume} {56}},\ \bibinfo
  {pages} {2449--2452} (\bibinfo {year} {1986})}\BibitemShut {NoStop}%
\bibitem [{\citenamefont {Wilkie}\ and\ \citenamefont
  {Brumer}(1991)}]{WilkieBrumer1991}%
  \BibitemOpen
  \bibfield  {author} {\bibinfo {author} {\bibfnamefont {Joshua}\ \bibnamefont
  {Wilkie}}\ and\ \bibinfo {author} {\bibfnamefont {Paul}\ \bibnamefont
  {Brumer}},\ }\bibfield  {title} {\enquote {\bibinfo {title} {Time-dependent
  manifestations of quantum chaos},}\ }\href {\doibase
  10.1103/PhysRevLett.67.1185} {\bibfield  {journal} {\bibinfo  {journal}
  {Phys. Rev. Lett.}\ }\textbf {\bibinfo {volume} {67}},\ \bibinfo {pages}
  {1185--1188} (\bibinfo {year} {1991})}\BibitemShut {NoStop}%
\bibitem [{\citenamefont {Br\'ezin}\ and\ \citenamefont
  {Hikami}(1997)}]{Brezin1997}%
  \BibitemOpen
  \bibfield  {author} {\bibinfo {author} {\bibfnamefont {E.}~\bibnamefont
  {Br\'ezin}}\ and\ \bibinfo {author} {\bibfnamefont {S.}~\bibnamefont
  {Hikami}},\ }\bibfield  {title} {\enquote {\bibinfo {title} {Spectral form
  factor in a random matrix theory},}\ }\href {\doibase
  10.1103/PhysRevE.55.4067} {\bibfield  {journal} {\bibinfo  {journal} {Phys.
  Rev. E}\ }\textbf {\bibinfo {volume} {55}},\ \bibinfo {pages} {4067--4083}
  (\bibinfo {year} {1997})}\BibitemShut {NoStop}%
\bibitem [{\citenamefont {Cotler}\ \emph {et~al.}(2017)\citenamefont {Cotler},
  \citenamefont {Gur-Ari}, \citenamefont {Hanada}, \citenamefont {Polchinski},
  \citenamefont {Saad}, \citenamefont {Shenker}, \citenamefont {Stanford},
  \citenamefont {Streicher},\ and\ \citenamefont {Tezuka}}]{Cotler2017}%
  \BibitemOpen
  \bibfield  {author} {\bibinfo {author} {\bibfnamefont {Jordan~S.}\
  \bibnamefont {Cotler}}, \bibinfo {author} {\bibfnamefont {Guy}\ \bibnamefont
  {Gur-Ari}}, \bibinfo {author} {\bibfnamefont {Masanori}\ \bibnamefont
  {Hanada}}, \bibinfo {author} {\bibfnamefont {Joseph}\ \bibnamefont
  {Polchinski}}, \bibinfo {author} {\bibfnamefont {Phil}\ \bibnamefont {Saad}},
  \bibinfo {author} {\bibfnamefont {Stephen~H.}\ \bibnamefont {Shenker}},
  \bibinfo {author} {\bibfnamefont {Douglas}\ \bibnamefont {Stanford}},
  \bibinfo {author} {\bibfnamefont {Alexandre}\ \bibnamefont {Streicher}}, \
  and\ \bibinfo {author} {\bibfnamefont {Masaki}\ \bibnamefont {Tezuka}},\
  }\bibfield  {title} {\enquote {\bibinfo {title} {{Black Holes and Random
  Matrices}},}\ }\href {\doibase 10.1007/JHEP05(2017)118} {\bibfield  {journal}
  {\bibinfo  {journal} {JHEP}\ }\textbf {\bibinfo {volume} {05}},\ \bibinfo
  {pages} {118} (\bibinfo {year} {2017})},\ \bibinfo {note} {[Erratum: JHEP 09,
  002 (2018)]}\BibitemShut {NoStop}%
\bibitem [{\citenamefont {Saad}\ \emph {et~al.}(2018)\citenamefont {Saad},
  \citenamefont {Shenker},\ and\ \citenamefont {Stanford}}]{Saad:2018bqo}%
  \BibitemOpen
  \bibfield  {author} {\bibinfo {author} {\bibfnamefont {Phil}\ \bibnamefont
  {Saad}}, \bibinfo {author} {\bibfnamefont {Stephen~H.}\ \bibnamefont
  {Shenker}}, \ and\ \bibinfo {author} {\bibfnamefont {Douglas}\ \bibnamefont
  {Stanford}},\ }\bibfield  {title} {\enquote {\bibinfo {title} {{A
  semiclassical ramp in SYK and in gravity}},}\ }\href@noop {} {\  (\bibinfo
  {year} {2018})},\ \Eprint {http://arxiv.org/abs/1806.06840} {arXiv:1806.06840
  [hep-th]} \BibitemShut {NoStop}%
\bibitem [{\citenamefont {Das}\ \emph {et~al.}(2024)\citenamefont {Das} \emph
  {et~al.}}]{das2024proposal}%
  \BibitemOpen
  \bibfield  {author} {\bibinfo {author} {\bibfnamefont {Adway~Kumar}\
  \bibnamefont {Das}} \emph {et~al.},\ }\bibfield  {title} {\enquote {\bibinfo
  {title} {{Proposal for many-body quantum chaos detection}},}\ }\href@noop {}
  {\  (\bibinfo {year} {2024})},\ \Eprint {http://arxiv.org/abs/2401.01401}
  {arXiv:2401.01401 [cond-mat.stat-mech]} \BibitemShut {NoStop}%
\bibitem [{\citenamefont {Dong}\ \emph {et~al.}(2024)\citenamefont {Dong} \emph
  {et~al.}}]{Dong:2024yaf}%
  \BibitemOpen
  \bibfield  {author} {\bibinfo {author} {\bibfnamefont {Hang}\ \bibnamefont
  {Dong}} \emph {et~al.},\ }\bibfield  {title} {\enquote {\bibinfo {title}
  {{Measuring Spectral Form Factor in Many-Body Chaotic and Localized Phases of
  Quantum Processors}},}\ }\href@noop {} {\  (\bibinfo {year} {2024})},\
  \Eprint {http://arxiv.org/abs/2403.16935} {arXiv:2403.16935 [quant-ph]}
  \BibitemShut {NoStop}%
\bibitem [{\citenamefont {Denisov}\ \emph {et~al.}(2019)\citenamefont
  {Denisov}, \citenamefont {Laptyeva}, \citenamefont {Tarnowski}, \citenamefont
  {Chru\'sci\'nski},\ and\ \citenamefont {\.Zyczkowski}}]{Denisov:2018nif}%
  \BibitemOpen
  \bibfield  {author} {\bibinfo {author} {\bibfnamefont {Sergey}\ \bibnamefont
  {Denisov}}, \bibinfo {author} {\bibfnamefont {Tetyana}\ \bibnamefont
  {Laptyeva}}, \bibinfo {author} {\bibfnamefont {Wojciech}\ \bibnamefont
  {Tarnowski}}, \bibinfo {author} {\bibfnamefont {Dariusz}\ \bibnamefont
  {Chru\'sci\'nski}}, \ and\ \bibinfo {author} {\bibfnamefont {Karol}\
  \bibnamefont {\.Zyczkowski}},\ }\bibfield  {title} {\enquote {\bibinfo
  {title} {{Universal spectra of random Lindblad operators}},}\ }\href
  {\doibase 10.1103/PhysRevLett.123.140403} {\bibfield  {journal} {\bibinfo
  {journal} {Phys. Rev. Lett.}\ }\textbf {\bibinfo {volume} {123}},\ \bibinfo
  {pages} {140403} (\bibinfo {year} {2019})}\BibitemShut {NoStop}%
\bibitem [{\citenamefont {Yang}\ \emph {et~al.}(2024)\citenamefont {Yang},
  \citenamefont {Xu},\ and\ \citenamefont {del Campo}}]{Yang:2024boo}%
  \BibitemOpen
  \bibfield  {author} {\bibinfo {author} {\bibfnamefont {Yifeng}\ \bibnamefont
  {Yang}}, \bibinfo {author} {\bibfnamefont {Zhenyu}\ \bibnamefont {Xu}}, \
  and\ \bibinfo {author} {\bibfnamefont {Adolfo}\ \bibnamefont {del Campo}},\
  }\bibfield  {title} {\enquote {\bibinfo {title} {{Decoherence rate in random
  Lindblad dynamics}},}\ }\href {\doibase 10.1103/PhysRevResearch.6.023229}
  {\bibfield  {journal} {\bibinfo  {journal} {Phys. Rev. Res.}\ }\textbf
  {\bibinfo {volume} {6}},\ \bibinfo {pages} {023229} (\bibinfo {year}
  {2024})}\BibitemShut {NoStop}%
\bibitem [{\citenamefont {Nandy}\ \emph {et~al.}(2024)\citenamefont {Nandy},
  \citenamefont {Matsoukas-Roubeas}, \citenamefont {Mart\'\i{}nez-Azcona},
  \citenamefont {Dymarsky},\ and\ \citenamefont {del Campo}}]{Nandy:2024htc}%
  \BibitemOpen
  \bibfield  {author} {\bibinfo {author} {\bibfnamefont {Pratik}\ \bibnamefont
  {Nandy}}, \bibinfo {author} {\bibfnamefont {Apollonas~S.}\ \bibnamefont
  {Matsoukas-Roubeas}}, \bibinfo {author} {\bibfnamefont {Pablo}\ \bibnamefont
  {Mart\'\i{}nez-Azcona}}, \bibinfo {author} {\bibfnamefont {Anatoly}\
  \bibnamefont {Dymarsky}}, \ and\ \bibinfo {author} {\bibfnamefont {Adolfo}\
  \bibnamefont {del Campo}},\ }\bibfield  {title} {\enquote {\bibinfo {title}
  {{Quantum Dynamics in Krylov Space: Methods and Applications}},}\ }\href@noop
  {} {\  (\bibinfo {year} {2024})},\ \Eprint {http://arxiv.org/abs/2405.09628}
  {arXiv:2405.09628 [quant-ph]} \BibitemShut {NoStop}%
\bibitem [{\citenamefont {Bender}\ and\ \citenamefont
  {Boettcher}(1998)}]{Bender:1998ke}%
  \BibitemOpen
  \bibfield  {author} {\bibinfo {author} {\bibfnamefont {Carl~M.}\ \bibnamefont
  {Bender}}\ and\ \bibinfo {author} {\bibfnamefont {Stefan}\ \bibnamefont
  {Boettcher}},\ }\bibfield  {title} {\enquote {\bibinfo {title} {{Real spectra
  in nonHermitian Hamiltonians having PT symmetry}},}\ }\href {\doibase
  10.1103/PhysRevLett.80.5243} {\bibfield  {journal} {\bibinfo  {journal}
  {Phys. Rev. Lett.}\ }\textbf {\bibinfo {volume} {80}},\ \bibinfo {pages}
  {5243--5246} (\bibinfo {year} {1998})}\BibitemShut {NoStop}%
\bibitem [{\citenamefont {Shibata}\ and\ \citenamefont
  {Katsura}(2019)}]{Katsurasissipative}%
  \BibitemOpen
  \bibfield  {author} {\bibinfo {author} {\bibfnamefont {Naoyuki}\ \bibnamefont
  {Shibata}}\ and\ \bibinfo {author} {\bibfnamefont {Hosho}\ \bibnamefont
  {Katsura}},\ }\bibfield  {title} {\enquote {\bibinfo {title} {{Dissipative
  spin chain as a non-Hermitian Kitaev ladder}},}\ }\href {\doibase
  10.1103/PhysRevB.99.174303} {\bibfield  {journal} {\bibinfo  {journal} {Phys.
  Rev. B}\ }\textbf {\bibinfo {volume} {99}},\ \bibinfo {pages} {174303}
  (\bibinfo {year} {2019})}\BibitemShut {NoStop}%
\bibitem [{\citenamefont {Ashida}\ \emph {et~al.}(2021)\citenamefont {Ashida},
  \citenamefont {Gong},\ and\ \citenamefont {Ueda}}]{Ashida:2020dkc}%
  \BibitemOpen
  \bibfield  {author} {\bibinfo {author} {\bibfnamefont {Yuto}\ \bibnamefont
  {Ashida}}, \bibinfo {author} {\bibfnamefont {Zongping}\ \bibnamefont {Gong}},
  \ and\ \bibinfo {author} {\bibfnamefont {Masahito}\ \bibnamefont {Ueda}},\
  }\bibfield  {title} {\enquote {\bibinfo {title} {{Non-Hermitian physics}},}\
  }\href {\doibase 10.1080/00018732.2021.1876991} {\bibfield  {journal}
  {\bibinfo  {journal} {Adv. Phys.}\ }\textbf {\bibinfo {volume} {69}},\
  \bibinfo {pages} {249--435} (\bibinfo {year} {2021})}\BibitemShut {NoStop}%
\bibitem [{\citenamefont {Cornelius}\ \emph {et~al.}(2022)\citenamefont
  {Cornelius}, \citenamefont {Xu}, \citenamefont {Saxena}, \citenamefont
  {Chenu},\ and\ \citenamefont {del Campo}}]{Cornelius:2021ccu}%
  \BibitemOpen
  \bibfield  {author} {\bibinfo {author} {\bibfnamefont {Julien}\ \bibnamefont
  {Cornelius}}, \bibinfo {author} {\bibfnamefont {Zhenyu}\ \bibnamefont {Xu}},
  \bibinfo {author} {\bibfnamefont {Avadh}\ \bibnamefont {Saxena}}, \bibinfo
  {author} {\bibfnamefont {Aurelia}\ \bibnamefont {Chenu}}, \ and\ \bibinfo
  {author} {\bibfnamefont {Adolfo}\ \bibnamefont {del Campo}},\ }\bibfield
  {title} {\enquote {\bibinfo {title} {{Spectral Filtering Induced by
  Non-Hermitian Evolution with Balanced Gain and Loss: Enhancing Quantum
  Chaos}},}\ }\href {\doibase 10.1103/PhysRevLett.128.190402} {\bibfield
  {journal} {\bibinfo  {journal} {Phys. Rev. Lett.}\ }\textbf {\bibinfo
  {volume} {128}},\ \bibinfo {pages} {190402} (\bibinfo {year}
  {2022})}\BibitemShut {NoStop}%
\bibitem [{\citenamefont {Matsoukas-Roubeas}\ \emph
  {et~al.}(2023{\natexlab{a}})\citenamefont {Matsoukas-Roubeas}, \citenamefont
  {Roccati}, \citenamefont {Cornelius}, \citenamefont {Xu}, \citenamefont
  {Chenu},\ and\ \citenamefont {del Campo}}]{Matsoukas-Roubeas:2022odk}%
  \BibitemOpen
  \bibfield  {author} {\bibinfo {author} {\bibfnamefont {Apollonas~S.}\
  \bibnamefont {Matsoukas-Roubeas}}, \bibinfo {author} {\bibfnamefont
  {Federico}\ \bibnamefont {Roccati}}, \bibinfo {author} {\bibfnamefont
  {Julien}\ \bibnamefont {Cornelius}}, \bibinfo {author} {\bibfnamefont
  {Zhenyu}\ \bibnamefont {Xu}}, \bibinfo {author} {\bibfnamefont {Aurelia}\
  \bibnamefont {Chenu}}, \ and\ \bibinfo {author} {\bibfnamefont {Adolfo}\
  \bibnamefont {del Campo}},\ }\bibfield  {title} {\enquote {\bibinfo {title}
  {{Non-Hermitian Hamiltonian deformations in quantum mechanics}},}\ }\href
  {\doibase 10.1007/JHEP01(2023)060} {\bibfield  {journal} {\bibinfo  {journal}
  {JHEP}\ }\textbf {\bibinfo {volume} {01}},\ \bibinfo {pages} {060} (\bibinfo
  {year} {2023}{\natexlab{a}})}\BibitemShut {NoStop}%
\bibitem [{\citenamefont {S\'a}\ \emph {et~al.}(2020)\citenamefont {S\'a},
  \citenamefont {Ribeiro},\ and\ \citenamefont
  {Prosen}}]{ComplexspacingProsen}%
  \BibitemOpen
  \bibfield  {author} {\bibinfo {author} {\bibfnamefont {Lucas}\ \bibnamefont
  {S\'a}}, \bibinfo {author} {\bibfnamefont {Pedro}\ \bibnamefont {Ribeiro}}, \
  and\ \bibinfo {author} {\bibfnamefont {Toma\v{z}}\ \bibnamefont {Prosen}},\
  }\bibfield  {title} {\enquote {\bibinfo {title} {Complex spacing ratios: A
  signature of dissipative quantum chaos},}\ }\href {\doibase
  10.1103/PhysRevX.10.021019} {\bibfield  {journal} {\bibinfo  {journal} {Phys.
  Rev. X}\ }\textbf {\bibinfo {volume} {10}},\ \bibinfo {pages} {021019}
  (\bibinfo {year} {2020})}\BibitemShut {NoStop}%
\bibitem [{\citenamefont {Li}\ \emph {et~al.}(2021)\citenamefont {Li},
  \citenamefont {Prosen},\ and\ \citenamefont {Chan}}]{ProsenDSFF}%
  \BibitemOpen
  \bibfield  {author} {\bibinfo {author} {\bibfnamefont {Jiachen}\ \bibnamefont
  {Li}}, \bibinfo {author} {\bibfnamefont {Toma\v{z}}\ \bibnamefont {Prosen}},
  \ and\ \bibinfo {author} {\bibfnamefont {Amos}\ \bibnamefont {Chan}},\
  }\bibfield  {title} {\enquote {\bibinfo {title} {{Spectral Statistics of
  Non-Hermitian Matrices and Dissipative Quantum Chaos}},}\ }\href {\doibase
  10.1103/PhysRevLett.127.170602} {\bibfield  {journal} {\bibinfo  {journal}
  {Phys. Rev. Lett.}\ }\textbf {\bibinfo {volume} {127}},\ \bibinfo {pages}
  {170602} (\bibinfo {year} {2021})}\BibitemShut {NoStop}%
\bibitem [{\citenamefont {Li}\ \emph {et~al.}(2024)\citenamefont {Li},
  \citenamefont {Yan}, \citenamefont {Prosen},\ and\ \citenamefont
  {Chan}}]{Li:2024uzg}%
  \BibitemOpen
  \bibfield  {author} {\bibinfo {author} {\bibfnamefont {Jiachen}\ \bibnamefont
  {Li}}, \bibinfo {author} {\bibfnamefont {Stephen}\ \bibnamefont {Yan}},
  \bibinfo {author} {\bibfnamefont {Toma\v{z}}\ \bibnamefont {Prosen}}, \ and\
  \bibinfo {author} {\bibfnamefont {Amos}\ \bibnamefont {Chan}},\ }\bibfield
  {title} {\enquote {\bibinfo {title} {{Spectral form factor in chaotic,
  localized, and integrable open quantum many-body systems}},}\ }\href@noop {}
  {\  (\bibinfo {year} {2024})},\ \Eprint {http://arxiv.org/abs/2405.01641}
  {arXiv:2405.01641 [cond-mat.stat-mech]} \BibitemShut {NoStop}%
\bibitem [{\citenamefont {Cipolloni}\ and\ \citenamefont
  {Grometto}(2024)}]{Cipolloni:2023mwh}%
  \BibitemOpen
  \bibfield  {author} {\bibinfo {author} {\bibfnamefont {Giorgio}\ \bibnamefont
  {Cipolloni}}\ and\ \bibinfo {author} {\bibfnamefont {Nicolo}\ \bibnamefont
  {Grometto}},\ }\bibfield  {title} {\enquote {\bibinfo {title} {{The
  Dissipative Spectral Form Factor for I.I.D. Matrices}},}\ }\href {\doibase
  10.1007/s10955-024-03237-4} {\bibfield  {journal} {\bibinfo  {journal} {J.
  Statist. Phys.}\ }\textbf {\bibinfo {volume} {191}},\ \bibinfo {pages} {21}
  (\bibinfo {year} {2024})}\BibitemShut {NoStop}%
\bibitem [{\citenamefont {Garc\'\i{}a-Garc\'\i{}a}\ \emph
  {et~al.}(2023{\natexlab{a}})\citenamefont {Garc\'\i{}a-Garc\'\i{}a},
  \citenamefont {S\'a},\ and\ \citenamefont {Verbaarschot}}]{GarcianonHSYK}%
  \BibitemOpen
  \bibfield  {author} {\bibinfo {author} {\bibfnamefont {Antonio~M.}\
  \bibnamefont {Garc\'\i{}a-Garc\'\i{}a}}, \bibinfo {author} {\bibfnamefont
  {Lucas}\ \bibnamefont {S\'a}}, \ and\ \bibinfo {author} {\bibfnamefont
  {Jacobus J.~M.}\ \bibnamefont {Verbaarschot}},\ }\bibfield  {title} {\enquote
  {\bibinfo {title} {{Universality and its limits in non-Hermitian many-body
  quantum chaos using the Sachdev-Ye-Kitaev model}},}\ }\href {\doibase
  10.1103/PhysRevD.107.066007} {\bibfield  {journal} {\bibinfo  {journal}
  {Phys. Rev. D}\ }\textbf {\bibinfo {volume} {107}},\ \bibinfo {pages}
  {066007} (\bibinfo {year} {2023}{\natexlab{a}})}\BibitemShut {NoStop}%
\bibitem [{\citenamefont {Xu}\ \emph {et~al.}(2021)\citenamefont {Xu},
  \citenamefont {Chenu}, \citenamefont {Prosen},\ and\ \citenamefont {del
  Campo}}]{Xu:2020wky}%
  \BibitemOpen
  \bibfield  {author} {\bibinfo {author} {\bibfnamefont {Zhenyu}\ \bibnamefont
  {Xu}}, \bibinfo {author} {\bibfnamefont {Aurelia}\ \bibnamefont {Chenu}},
  \bibinfo {author} {\bibfnamefont {Toma\v{z}}\ \bibnamefont {Prosen}}, \ and\
  \bibinfo {author} {\bibfnamefont {Adolfo}\ \bibnamefont {del Campo}},\
  }\bibfield  {title} {\enquote {\bibinfo {title} {{Thermofield dynamics:
  Quantum Chaos versus Decoherence}},}\ }\href {\doibase
  10.1103/PhysRevB.103.064309} {\bibfield  {journal} {\bibinfo  {journal}
  {Phys. Rev. B}\ }\textbf {\bibinfo {volume} {103}},\ \bibinfo {pages}
  {064309} (\bibinfo {year} {2021})}\BibitemShut {NoStop}%
\bibitem [{\citenamefont {Zhou}\ \emph {et~al.}(2024)\citenamefont {Zhou},
  \citenamefont {Zhou},\ and\ \citenamefont {Zhang}}]{Zhou:2023qmk}%
  \BibitemOpen
  \bibfield  {author} {\bibinfo {author} {\bibfnamefont {Yi-Neng}\ \bibnamefont
  {Zhou}}, \bibinfo {author} {\bibfnamefont {Tian-Gang}\ \bibnamefont {Zhou}},
  \ and\ \bibinfo {author} {\bibfnamefont {Pengfei}\ \bibnamefont {Zhang}},\
  }\bibfield  {title} {\enquote {\bibinfo {title} {{General properties of the
  spectral form factor in open quantum systems}},}\ }\href {\doibase
  10.1007/s11467-024-1406-7} {\bibfield  {journal} {\bibinfo  {journal} {Front.
  Phys. (Beijing)}\ }\textbf {\bibinfo {volume} {19}},\ \bibinfo {pages}
  {31202} (\bibinfo {year} {2024})}\BibitemShut {NoStop}%
\bibitem [{\citenamefont {Kawabata}\ \emph {et~al.}(2023)\citenamefont
  {Kawabata}, \citenamefont {Xiao}, \citenamefont {Ohtsuki},\ and\
  \citenamefont {Shindou}}]{KawabataSVD23}%
  \BibitemOpen
  \bibfield  {author} {\bibinfo {author} {\bibfnamefont {Kohei}\ \bibnamefont
  {Kawabata}}, \bibinfo {author} {\bibfnamefont {Zhenyu}\ \bibnamefont {Xiao}},
  \bibinfo {author} {\bibfnamefont {Tomi}\ \bibnamefont {Ohtsuki}}, \ and\
  \bibinfo {author} {\bibfnamefont {Ryuichi}\ \bibnamefont {Shindou}},\
  }\bibfield  {title} {\enquote {\bibinfo {title} {{Singular-Value Statistics
  of Non-Hermitian Random Matrices and Open Quantum Systems}},}\ }\href
  {\doibase 10.1103/PRXQuantum.4.040312} {\bibfield  {journal} {\bibinfo
  {journal} {PRX Quantum}\ }\textbf {\bibinfo {volume} {4}},\ \bibinfo {pages}
  {040312} (\bibinfo {year} {2023})}\BibitemShut {NoStop}%
\bibitem [{\citenamefont {Roccati}\ \emph {et~al.}(2024)\citenamefont
  {Roccati}, \citenamefont {Balducci}, \citenamefont {Shir},\ and\
  \citenamefont {Chenu}}]{ChenuSVD}%
  \BibitemOpen
  \bibfield  {author} {\bibinfo {author} {\bibfnamefont {Federico}\
  \bibnamefont {Roccati}}, \bibinfo {author} {\bibfnamefont {Federico}\
  \bibnamefont {Balducci}}, \bibinfo {author} {\bibfnamefont {Ruth}\
  \bibnamefont {Shir}}, \ and\ \bibinfo {author} {\bibfnamefont {Aur\'elia}\
  \bibnamefont {Chenu}},\ }\bibfield  {title} {\enquote {\bibinfo {title}
  {{Diagnosing non-Hermitian many-body localization and quantum chaos via
  singular value decomposition}},}\ }\href {\doibase
  10.1103/PhysRevB.109.L140201} {\bibfield  {journal} {\bibinfo  {journal}
  {Phys. Rev. B}\ }\textbf {\bibinfo {volume} {109}},\ \bibinfo {pages}
  {L140201} (\bibinfo {year} {2024})}\BibitemShut {NoStop}%
\bibitem [{\citenamefont {Hamazaki}\ \emph {et~al.}(2019)\citenamefont
  {Hamazaki}, \citenamefont {Kawabata},\ and\ \citenamefont
  {Ueda}}]{HamazakiKawabataUeda}%
  \BibitemOpen
  \bibfield  {author} {\bibinfo {author} {\bibfnamefont {Ryusuke}\ \bibnamefont
  {Hamazaki}}, \bibinfo {author} {\bibfnamefont {Kohei}\ \bibnamefont
  {Kawabata}}, \ and\ \bibinfo {author} {\bibfnamefont {Masahito}\ \bibnamefont
  {Ueda}},\ }\bibfield  {title} {\enquote {\bibinfo {title} {Non-hermitian
  many-body localization},}\ }\href {\doibase 10.1103/PhysRevLett.123.090603}
  {\bibfield  {journal} {\bibinfo  {journal} {Phys. Rev. Lett.}\ }\textbf
  {\bibinfo {volume} {123}},\ \bibinfo {pages} {090603} (\bibinfo {year}
  {2019})}\BibitemShut {NoStop}%
\bibitem [{\citenamefont {Martinez~Alvarez}\ \emph {et~al.}(2018)\citenamefont
  {Martinez~Alvarez}, \citenamefont {Barrios~Vargas},\ and\ \citenamefont
  {Foa~Torres}}]{nonHskin1}%
  \BibitemOpen
  \bibfield  {author} {\bibinfo {author} {\bibfnamefont {V.~M.}\ \bibnamefont
  {Martinez~Alvarez}}, \bibinfo {author} {\bibfnamefont {J.~E.}\ \bibnamefont
  {Barrios~Vargas}}, \ and\ \bibinfo {author} {\bibfnamefont {L.~E.~F.}\
  \bibnamefont {Foa~Torres}},\ }\bibfield  {title} {\enquote {\bibinfo {title}
  {Non-hermitian robust edge states in one dimension: Anomalous localization
  and eigenspace condensation at exceptional points},}\ }\href {\doibase
  10.1103/PhysRevB.97.121401} {\bibfield  {journal} {\bibinfo  {journal} {Phys.
  Rev. B}\ }\textbf {\bibinfo {volume} {97}},\ \bibinfo {pages} {121401(R)}
  (\bibinfo {year} {2018})}\BibitemShut {NoStop}%
\bibitem [{\citenamefont {Yao}\ and\ \citenamefont {Wang}(2018)}]{nonHskin2}%
  \BibitemOpen
  \bibfield  {author} {\bibinfo {author} {\bibfnamefont {Shunyu}\ \bibnamefont
  {Yao}}\ and\ \bibinfo {author} {\bibfnamefont {Zhong}\ \bibnamefont {Wang}},\
  }\bibfield  {title} {\enquote {\bibinfo {title} {Edge states and topological
  invariants of non-hermitian systems},}\ }\href {\doibase
  10.1103/PhysRevLett.121.086803} {\bibfield  {journal} {\bibinfo  {journal}
  {Phys. Rev. Lett.}\ }\textbf {\bibinfo {volume} {121}},\ \bibinfo {pages}
  {086803} (\bibinfo {year} {2018})}\BibitemShut {NoStop}%
\bibitem [{\citenamefont {Hamanaka}\ and\ \citenamefont
  {Kawabata}(2024)}]{Hamanaka:2024njv}%
  \BibitemOpen
  \bibfield  {author} {\bibinfo {author} {\bibfnamefont {Shu}\ \bibnamefont
  {Hamanaka}}\ and\ \bibinfo {author} {\bibfnamefont {Kohei}\ \bibnamefont
  {Kawabata}},\ }\bibfield  {title} {\enquote {\bibinfo {title}
  {{Multifractality of Many-Body Non-Hermitian Skin Effect}},}\ }\href@noop {}
  {\  (\bibinfo {year} {2024})},\ \Eprint {http://arxiv.org/abs/2401.08304}
  {arXiv:2401.08304 [cond-mat.str-el]} \BibitemShut {NoStop}%
\bibitem [{\citenamefont {Parzygnat}\ \emph {et~al.}(2023)\citenamefont
  {Parzygnat}, \citenamefont {Takayanagi}, \citenamefont {Taki},\ and\
  \citenamefont {Wei}}]{Parzygnat:2023avh}%
  \BibitemOpen
  \bibfield  {author} {\bibinfo {author} {\bibfnamefont {Arthur~J.}\
  \bibnamefont {Parzygnat}}, \bibinfo {author} {\bibfnamefont {Tadashi}\
  \bibnamefont {Takayanagi}}, \bibinfo {author} {\bibfnamefont {Yusuke}\
  \bibnamefont {Taki}}, \ and\ \bibinfo {author} {\bibfnamefont {Zixia}\
  \bibnamefont {Wei}},\ }\bibfield  {title} {\enquote {\bibinfo {title} {{SVD
  entanglement entropy}},}\ }\href {\doibase 10.1007/JHEP12(2023)123}
  {\bibfield  {journal} {\bibinfo  {journal} {JHEP}\ }\textbf {\bibinfo
  {volume} {12}},\ \bibinfo {pages} {123} (\bibinfo {year} {2023})}\BibitemShut
  {NoStop}%
\bibitem [{\citenamefont {Sachdev}\ and\ \citenamefont
  {Ye}(1993)}]{PhysRevLett.70.3339}%
  \BibitemOpen
  \bibfield  {author} {\bibinfo {author} {\bibfnamefont {Subir}\ \bibnamefont
  {Sachdev}}\ and\ \bibinfo {author} {\bibfnamefont {Jinwu}\ \bibnamefont
  {Ye}},\ }\bibfield  {title} {\enquote {\bibinfo {title} {Gapless spin-fluid
  ground state in a random quantum heisenberg magnet},}\ }\href {\doibase
  10.1103/PhysRevLett.70.3339} {\bibfield  {journal} {\bibinfo  {journal}
  {Phys. Rev. Lett.}\ }\textbf {\bibinfo {volume} {70}},\ \bibinfo {pages}
  {3339--3342} (\bibinfo {year} {1993})}\BibitemShut {NoStop}%
\bibitem [{\citenamefont {Kitaev}(2015)}]{Kittu}%
  \BibitemOpen
  \bibfield  {author} {\bibinfo {author} {\bibfnamefont {A.}~\bibnamefont
  {Kitaev}},\ }\href@noop {} {\enquote {\bibinfo {title} {A simple model of
  quantum holography (part 1) and (part 2)},}\ }\bibinfo {howpublished}
  {\url{https://online.kitp.ucsb.edu/online/joint98/kitaev/},
  \url{https://online.kitp.ucsb.edu/online/entangled15/kitaev2/}} (\bibinfo
  {year} {2015}),\ \bibinfo {note} {talk given at KITP}\BibitemShut {NoStop}%
\bibitem [{\citenamefont {Maldacena}\ and\ \citenamefont
  {Stanford}(2016)}]{Maldacena:2016hyu}%
  \BibitemOpen
  \bibfield  {author} {\bibinfo {author} {\bibfnamefont {Juan}\ \bibnamefont
  {Maldacena}}\ and\ \bibinfo {author} {\bibfnamefont {Douglas}\ \bibnamefont
  {Stanford}},\ }\bibfield  {title} {\enquote {\bibinfo {title} {{Remarks on
  the Sachdev-Ye-Kitaev model}},}\ }\href {\doibase 10.1103/PhysRevD.94.106002}
  {\bibfield  {journal} {\bibinfo  {journal} {Phys. Rev. D}\ }\textbf {\bibinfo
  {volume} {94}},\ \bibinfo {pages} {106002} (\bibinfo {year}
  {2016})}\BibitemShut {NoStop}%
\bibitem [{\citenamefont {Chowdhury}\ \emph {et~al.}(2022)\citenamefont
  {Chowdhury}, \citenamefont {Georges}, \citenamefont {Parcollet},\ and\
  \citenamefont {Sachdev}}]{RevModPhys.94.035004}%
  \BibitemOpen
  \bibfield  {author} {\bibinfo {author} {\bibfnamefont {Debanjan}\
  \bibnamefont {Chowdhury}}, \bibinfo {author} {\bibfnamefont {Antoine}\
  \bibnamefont {Georges}}, \bibinfo {author} {\bibfnamefont {Olivier}\
  \bibnamefont {Parcollet}}, \ and\ \bibinfo {author} {\bibfnamefont {Subir}\
  \bibnamefont {Sachdev}},\ }\bibfield  {title} {\enquote {\bibinfo {title}
  {{Sachdev-Ye-Kitaev models and beyond: Window into non-Fermi liquids}},}\
  }\href {\doibase 10.1103/RevModPhys.94.035004} {\bibfield  {journal}
  {\bibinfo  {journal} {Rev. Mod. Phys.}\ }\textbf {\bibinfo {volume} {94}},\
  \bibinfo {pages} {035004} (\bibinfo {year} {2022})}\BibitemShut {NoStop}%
\bibitem [{\citenamefont {Rosenhaus}(2019)}]{Rosenhaus:2018dtp}%
  \BibitemOpen
  \bibfield  {author} {\bibinfo {author} {\bibfnamefont {Vladimir}\
  \bibnamefont {Rosenhaus}},\ }\bibfield  {title} {\enquote {\bibinfo {title}
  {{An introduction to the SYK model}},}\ }\href {\doibase
  10.1088/1751-8121/ab2ce1} {\bibfield  {journal} {\bibinfo  {journal} {J.
  Phys. A}\ }\textbf {\bibinfo {volume} {52}},\ \bibinfo {pages} {323001}
  (\bibinfo {year} {2019})}\BibitemShut {NoStop}%
\bibitem [{\citenamefont {Xu}\ \emph {et~al.}(2020)\citenamefont {Xu},
  \citenamefont {Susskind}, \citenamefont {Su},\ and\ \citenamefont
  {Swingle}}]{Xu:2020shn}%
  \BibitemOpen
  \bibfield  {author} {\bibinfo {author} {\bibfnamefont {Shenglong}\
  \bibnamefont {Xu}}, \bibinfo {author} {\bibfnamefont {Leonard}\ \bibnamefont
  {Susskind}}, \bibinfo {author} {\bibfnamefont {Yuan}\ \bibnamefont {Su}}, \
  and\ \bibinfo {author} {\bibfnamefont {Brian}\ \bibnamefont {Swingle}},\
  }\bibfield  {title} {\enquote {\bibinfo {title} {{A Sparse Model of Quantum
  Holography}},}\ }\href@noop {} {\  (\bibinfo {year} {2020})},\ \Eprint
  {http://arxiv.org/abs/2008.02303} {arXiv:2008.02303 [cond-mat.str-el]}
  \BibitemShut {NoStop}%
\bibitem [{\citenamefont {Tezuka}\ \emph {et~al.}(2023)\citenamefont {Tezuka},
  \citenamefont {Oktay}, \citenamefont {Rinaldi}, \citenamefont {Hanada},\ and\
  \citenamefont {Nori}}]{Tezuka:2022mrr}%
  \BibitemOpen
  \bibfield  {author} {\bibinfo {author} {\bibfnamefont {Masaki}\ \bibnamefont
  {Tezuka}}, \bibinfo {author} {\bibfnamefont {Onur}\ \bibnamefont {Oktay}},
  \bibinfo {author} {\bibfnamefont {Enrico}\ \bibnamefont {Rinaldi}}, \bibinfo
  {author} {\bibfnamefont {Masanori}\ \bibnamefont {Hanada}}, \ and\ \bibinfo
  {author} {\bibfnamefont {Franco}\ \bibnamefont {Nori}},\ }\bibfield  {title}
  {\enquote {\bibinfo {title} {{Binary-coupling sparse Sachdev-Ye-Kitaev model:
  An improved model of quantum chaos and holography}},}\ }\href {\doibase
  10.1103/PhysRevB.107.L081103} {\bibfield  {journal} {\bibinfo  {journal}
  {Phys. Rev. B}\ }\textbf {\bibinfo {volume} {107}},\ \bibinfo {pages}
  {L081103} (\bibinfo {year} {2023})}\BibitemShut {NoStop}%
\bibitem [{\citenamefont {Anegawa}\ \emph {et~al.}(2023)\citenamefont
  {Anegawa}, \citenamefont {Iizuka}, \citenamefont {Mukherjee}, \citenamefont
  {Sake},\ and\ \citenamefont {Trivedi}}]{Anegawa:2023vxq}%
  \BibitemOpen
  \bibfield  {author} {\bibinfo {author} {\bibfnamefont {Takanori}\
  \bibnamefont {Anegawa}}, \bibinfo {author} {\bibfnamefont {Norihiro}\
  \bibnamefont {Iizuka}}, \bibinfo {author} {\bibfnamefont {Arkaprava}\
  \bibnamefont {Mukherjee}}, \bibinfo {author} {\bibfnamefont {Sunil~Kumar}\
  \bibnamefont {Sake}}, \ and\ \bibinfo {author} {\bibfnamefont {Sandip~P.}\
  \bibnamefont {Trivedi}},\ }\bibfield  {title} {\enquote {\bibinfo {title}
  {{Sparse random matrices and Gaussian ensembles with varying randomness}},}\
  }\href {\doibase 10.1007/JHEP11(2023)234} {\bibfield  {journal} {\bibinfo
  {journal} {JHEP}\ }\textbf {\bibinfo {volume} {11}},\ \bibinfo {pages} {234}
  (\bibinfo {year} {2023})}\BibitemShut {NoStop}%
\bibitem [{\citenamefont {Caceres}\ \emph {et~al.}(2021)\citenamefont
  {Caceres}, \citenamefont {Misobuchi},\ and\ \citenamefont
  {Pimentel}}]{Caceres:2021nsa}%
  \BibitemOpen
  \bibfield  {author} {\bibinfo {author} {\bibfnamefont {Elena}\ \bibnamefont
  {Caceres}}, \bibinfo {author} {\bibfnamefont {Anderson}\ \bibnamefont
  {Misobuchi}}, \ and\ \bibinfo {author} {\bibfnamefont {Rafael}\ \bibnamefont
  {Pimentel}},\ }\bibfield  {title} {\enquote {\bibinfo {title} {{Sparse SYK
  and traversable wormholes}},}\ }\href {\doibase 10.1007/JHEP11(2021)015}
  {\bibfield  {journal} {\bibinfo  {journal} {JHEP}\ }\textbf {\bibinfo
  {volume} {11}},\ \bibinfo {pages} {015} (\bibinfo {year} {2021})}\BibitemShut
  {NoStop}%
\bibitem [{\citenamefont {C\'aceres}\ \emph {et~al.}(2023)\citenamefont
  {C\'aceres}, \citenamefont {Guglielmo}, \citenamefont {Kent},\ and\
  \citenamefont {Misobuchi}}]{Caceres:2023yoj}%
  \BibitemOpen
  \bibfield  {author} {\bibinfo {author} {\bibfnamefont {Elena}\ \bibnamefont
  {C\'aceres}}, \bibinfo {author} {\bibfnamefont {Tyler}\ \bibnamefont
  {Guglielmo}}, \bibinfo {author} {\bibfnamefont {Brian}\ \bibnamefont {Kent}},
  \ and\ \bibinfo {author} {\bibfnamefont {Anderson}\ \bibnamefont
  {Misobuchi}},\ }\bibfield  {title} {\enquote {\bibinfo {title}
  {{Out-of-time-order correlators and Lyapunov exponents in sparse SYK}},}\
  }\href {\doibase 10.1007/JHEP11(2023)088} {\bibfield  {journal} {\bibinfo
  {journal} {JHEP}\ }\textbf {\bibinfo {volume} {11}},\ \bibinfo {pages} {088}
  (\bibinfo {year} {2023})}\BibitemShut {NoStop}%
\bibitem [{\citenamefont {Garc\'\i{}a-Garc\'\i{}a}\ \emph
  {et~al.}(2021)\citenamefont {Garc\'\i{}a-Garc\'\i{}a}, \citenamefont {Jia},
  \citenamefont {Rosa},\ and\ \citenamefont
  {Verbaarschot}}]{Garcia-Garcia:2020cdo}%
  \BibitemOpen
  \bibfield  {author} {\bibinfo {author} {\bibfnamefont {Antonio~M.}\
  \bibnamefont {Garc\'\i{}a-Garc\'\i{}a}}, \bibinfo {author} {\bibfnamefont
  {Yiyang}\ \bibnamefont {Jia}}, \bibinfo {author} {\bibfnamefont {Dario}\
  \bibnamefont {Rosa}}, \ and\ \bibinfo {author} {\bibfnamefont {Jacobus
  J.~M.}\ \bibnamefont {Verbaarschot}},\ }\bibfield  {title} {\enquote
  {\bibinfo {title} {{Sparse Sachdev-Ye-Kitaev model, quantum chaos and gravity
  duals}},}\ }\href {\doibase 10.1103/PhysRevD.103.106002} {\bibfield
  {journal} {\bibinfo  {journal} {Phys. Rev. D}\ }\textbf {\bibinfo {volume}
  {103}},\ \bibinfo {pages} {106002} (\bibinfo {year} {2021})}\BibitemShut
  {NoStop}%
\bibitem [{\citenamefont {Orman}\ \emph {et~al.}(2024)\citenamefont {Orman},
  \citenamefont {Gharibyan},\ and\ \citenamefont {Preskill}}]{Orman:2024mpw}%
  \BibitemOpen
  \bibfield  {author} {\bibinfo {author} {\bibfnamefont {Patrick}\ \bibnamefont
  {Orman}}, \bibinfo {author} {\bibfnamefont {Hrant}\ \bibnamefont
  {Gharibyan}}, \ and\ \bibinfo {author} {\bibfnamefont {John}\ \bibnamefont
  {Preskill}},\ }\bibfield  {title} {\enquote {\bibinfo {title} {{Quantum chaos
  in the sparse SYK model}},}\ }\href@noop {} {\  (\bibinfo {year} {2024})},\
  \Eprint {http://arxiv.org/abs/2403.13884} {arXiv:2403.13884 [hep-th]}
  \BibitemShut {NoStop}%
\bibitem [{\citenamefont {Jafferis}\ \emph {et~al.}(2022)\citenamefont
  {Jafferis}, \citenamefont {Zlokapa}, \citenamefont {Lykken}, \citenamefont
  {Kolchmeyer}, \citenamefont {Davis}, \citenamefont {Lauk}, \citenamefont
  {Neven},\ and\ \citenamefont {Spiropulu}}]{Jafferis2022}%
  \BibitemOpen
  \bibfield  {author} {\bibinfo {author} {\bibfnamefont {Daniel}\ \bibnamefont
  {Jafferis}}, \bibinfo {author} {\bibfnamefont {Alexander}\ \bibnamefont
  {Zlokapa}}, \bibinfo {author} {\bibfnamefont {Joseph~D.}\ \bibnamefont
  {Lykken}}, \bibinfo {author} {\bibfnamefont {David~K.}\ \bibnamefont
  {Kolchmeyer}}, \bibinfo {author} {\bibfnamefont {Samantha~I.}\ \bibnamefont
  {Davis}}, \bibinfo {author} {\bibfnamefont {Nikolai}\ \bibnamefont {Lauk}},
  \bibinfo {author} {\bibfnamefont {Hartmut}\ \bibnamefont {Neven}}, \ and\
  \bibinfo {author} {\bibfnamefont {Maria}\ \bibnamefont {Spiropulu}},\
  }\bibfield  {title} {\enquote {\bibinfo {title} {Traversable wormhole
  dynamics on a quantum processor},}\ }\href {\doibase
  10.1038/s41586-022-05424-3} {\bibfield  {journal} {\bibinfo  {journal}
  {Nature}\ }\textbf {\bibinfo {volume} {612}},\ \bibinfo {pages} {51--55}
  (\bibinfo {year} {2022})}\BibitemShut {NoStop}%
\bibitem [{\citenamefont {Kobrin}\ \emph {et~al.}(2023)\citenamefont {Kobrin},
  \citenamefont {Schuster},\ and\ \citenamefont {Yao}}]{Kobrin:2023rzr}%
  \BibitemOpen
  \bibfield  {author} {\bibinfo {author} {\bibfnamefont {Bryce}\ \bibnamefont
  {Kobrin}}, \bibinfo {author} {\bibfnamefont {Thomas}\ \bibnamefont
  {Schuster}}, \ and\ \bibinfo {author} {\bibfnamefont {Norman~Y.}\
  \bibnamefont {Yao}},\ }\bibfield  {title} {\enquote {\bibinfo {title}
  {{Comment on `Traversable wormhole dynamics on a quantum processor'}},}\
  }\href@noop {} {\  (\bibinfo {year} {2023})},\ \Eprint
  {http://arxiv.org/abs/2302.07897} {arXiv:2302.07897 [quant-ph]} \BibitemShut
  {NoStop}%
\bibitem [{\citenamefont {Iliesiu}\ \emph {et~al.}(2022)\citenamefont
  {Iliesiu}, \citenamefont {Mezei},\ and\ \citenamefont
  {S\'arosi}}]{Iliesiu:2021ari}%
  \BibitemOpen
  \bibfield  {author} {\bibinfo {author} {\bibfnamefont {Luca~V.}\ \bibnamefont
  {Iliesiu}}, \bibinfo {author} {\bibfnamefont {M\'ark}\ \bibnamefont {Mezei}},
  \ and\ \bibinfo {author} {\bibfnamefont {G\'abor}\ \bibnamefont {S\'arosi}},\
  }\bibfield  {title} {\enquote {\bibinfo {title} {{The volume of the black
  hole interior at late times}},}\ }\href {\doibase 10.1007/JHEP07(2022)073}
  {\bibfield  {journal} {\bibinfo  {journal} {JHEP}\ }\textbf {\bibinfo
  {volume} {07}},\ \bibinfo {pages} {073} (\bibinfo {year} {2022})}\BibitemShut
  {NoStop}%
\bibitem [{\citenamefont {Lau}\ \emph {et~al.}(2021)\citenamefont {Lau},
  \citenamefont {Ma}, \citenamefont {Murugan},\ and\ \citenamefont
  {Tezuka}}]{Lau:2020qnl}%
  \BibitemOpen
  \bibfield  {author} {\bibinfo {author} {\bibfnamefont {Pak Hang~Chris}\
  \bibnamefont {Lau}}, \bibinfo {author} {\bibfnamefont {Chen-Te}\ \bibnamefont
  {Ma}}, \bibinfo {author} {\bibfnamefont {Jeff}\ \bibnamefont {Murugan}}, \
  and\ \bibinfo {author} {\bibfnamefont {Masaki}\ \bibnamefont {Tezuka}},\
  }\bibfield  {title} {\enquote {\bibinfo {title} {{Correlated disorder in the
  SYK$_2$ model}},}\ }\href {\doibase 10.1088/1751-8121/abde77} {\bibfield
  {journal} {\bibinfo  {journal} {J. Phys. A}\ }\textbf {\bibinfo {volume}
  {54}},\ \bibinfo {pages} {095401} (\bibinfo {year} {2021})}\BibitemShut
  {NoStop}%
\bibitem [{\citenamefont {Ozaki}\ and\ \citenamefont
  {Katsura}(2025)}]{Ozaki:2024wpj}%
  \BibitemOpen
  \bibfield  {author} {\bibinfo {author} {\bibfnamefont {Soshun}\ \bibnamefont
  {Ozaki}}\ and\ \bibinfo {author} {\bibfnamefont {Hosho}\ \bibnamefont
  {Katsura}},\ }\bibfield  {title} {\enquote {\bibinfo {title} {{Disorder-free
  Sachdev-Ye-Kitaev models: Integrability and a precursor of chaos}},}\ }\href
  {\doibase 10.1103/PhysRevResearch.7.013092} {\bibfield  {journal} {\bibinfo
  {journal} {Phys. Rev. Res.}\ }\textbf {\bibinfo {volume} {7}},\ \bibinfo
  {pages} {013092} (\bibinfo {year} {2025})}\BibitemShut {NoStop}%
\bibitem [{\citenamefont {Garc\'\i{}a-Garc\'\i{}a}\ and\ \citenamefont
  {Godet}(2021)}]{Garcia-Garcia:2020ttf}%
  \BibitemOpen
  \bibfield  {author} {\bibinfo {author} {\bibfnamefont {Antonio~M.}\
  \bibnamefont {Garc\'\i{}a-Garc\'\i{}a}}\ and\ \bibinfo {author}
  {\bibfnamefont {Victor}\ \bibnamefont {Godet}},\ }\bibfield  {title}
  {\enquote {\bibinfo {title} {{Euclidean wormhole in the Sachdev-Ye-Kitaev
  model}},}\ }\href {\doibase 10.1103/PhysRevD.103.046014} {\bibfield
  {journal} {\bibinfo  {journal} {Phys. Rev. D}\ }\textbf {\bibinfo {volume}
  {103}},\ \bibinfo {pages} {046014} (\bibinfo {year} {2021})}\BibitemShut
  {NoStop}%
\bibitem [{\citenamefont {Garc\'\i{}a-Garc\'\i{}a}\ \emph
  {et~al.}(2022{\natexlab{a}})\citenamefont {Garc\'\i{}a-Garc\'\i{}a},
  \citenamefont {S\'a},\ and\ \citenamefont
  {Verbaarschot}}]{Garcia-Garcia:2021rle}%
  \BibitemOpen
  \bibfield  {author} {\bibinfo {author} {\bibfnamefont {Antonio~M.}\
  \bibnamefont {Garc\'\i{}a-Garc\'\i{}a}}, \bibinfo {author} {\bibfnamefont
  {Lucas}\ \bibnamefont {S\'a}}, \ and\ \bibinfo {author} {\bibfnamefont
  {Jacobus J.~M.}\ \bibnamefont {Verbaarschot}},\ }\bibfield  {title} {\enquote
  {\bibinfo {title} {{Symmetry Classification and Universality in Non-Hermitian
  Many-Body Quantum Chaos by the Sachdev-Ye-Kitaev Model}},}\ }\href {\doibase
  10.1103/PhysRevX.12.021040} {\bibfield  {journal} {\bibinfo  {journal} {Phys.
  Rev. X}\ }\textbf {\bibinfo {volume} {12}},\ \bibinfo {pages} {021040}
  (\bibinfo {year} {2022}{\natexlab{a}})}\BibitemShut {NoStop}%
\bibitem [{\citenamefont {Garc\'\i{}a-Garc\'\i{}a}\ \emph
  {et~al.}(2022{\natexlab{b}})\citenamefont {Garc\'\i{}a-Garc\'\i{}a},
  \citenamefont {Jia}, \citenamefont {Rosa},\ and\ \citenamefont
  {Verbaarschot}}]{Garcia-Garcia:2022xsh}%
  \BibitemOpen
  \bibfield  {author} {\bibinfo {author} {\bibfnamefont {Antonio~M.}\
  \bibnamefont {Garc\'\i{}a-Garc\'\i{}a}}, \bibinfo {author} {\bibfnamefont
  {Yiyang}\ \bibnamefont {Jia}}, \bibinfo {author} {\bibfnamefont {Dario}\
  \bibnamefont {Rosa}}, \ and\ \bibinfo {author} {\bibfnamefont {Jacobus
  J.~M.}\ \bibnamefont {Verbaarschot}},\ }\bibfield  {title} {\enquote
  {\bibinfo {title} {{Replica symmetry breaking in random non-Hermitian
  systems}},}\ }\href {\doibase 10.1103/PhysRevD.105.126027} {\bibfield
  {journal} {\bibinfo  {journal} {Phys. Rev. D}\ }\textbf {\bibinfo {volume}
  {105}},\ \bibinfo {pages} {126027} (\bibinfo {year}
  {2022}{\natexlab{b}})}\BibitemShut {NoStop}%
\bibitem [{\citenamefont {Cipolloni}\ and\ \citenamefont
  {Kudler-Flam}(2023)}]{Cipolloni:2022fej}%
  \BibitemOpen
  \bibfield  {author} {\bibinfo {author} {\bibfnamefont {Giorgio}\ \bibnamefont
  {Cipolloni}}\ and\ \bibinfo {author} {\bibfnamefont {Jonah}\ \bibnamefont
  {Kudler-Flam}},\ }\bibfield  {title} {\enquote {\bibinfo {title}
  {{Entanglement Entropy of Non-Hermitian Eigenstates and the Ginibre
  Ensemble}},}\ }\href {\doibase 10.1103/PhysRevLett.130.010401} {\bibfield
  {journal} {\bibinfo  {journal} {Phys. Rev. Lett.}\ }\textbf {\bibinfo
  {volume} {130}},\ \bibinfo {pages} {010401} (\bibinfo {year}
  {2023})}\BibitemShut {NoStop}%
\bibitem [{\citenamefont {Nielsen}\ and\ \citenamefont
  {Chuang}(2010)}]{Nielsen_Chuang_2010}%
  \BibitemOpen
  \bibfield  {author} {\bibinfo {author} {\bibfnamefont {Michael~A.}\
  \bibnamefont {Nielsen}}\ and\ \bibinfo {author} {\bibfnamefont {Isaac~L.}\
  \bibnamefont {Chuang}},\ }\href {\doibase
  https://doi.org/10.1017/CBO9780511976667} {\emph {\bibinfo {title} {Quantum
  Computation and Quantum Information: 10th Anniversary Edition}}}\ (\bibinfo
  {publisher} {Cambridge University Press},\ \bibinfo {year}
  {2010})\BibitemShut {NoStop}%
\bibitem [{\citenamefont {Allard}\ and\ \citenamefont
  {Kieburg}(2024)}]{allard2024correlation}%
  \BibitemOpen
  \bibfield  {author} {\bibinfo {author} {\bibfnamefont {Matthias}\
  \bibnamefont {Allard}}\ and\ \bibinfo {author} {\bibfnamefont {Mario}\
  \bibnamefont {Kieburg}},\ }\bibfield  {title} {\enquote {\bibinfo {title}
  {Correlation functions between singular values and eigenvalues},}\
  }\href@noop {} {\  (\bibinfo {year} {2024})},\ \Eprint
  {http://arxiv.org/abs/2403.19157} {arXiv:2403.19157 [math.PR]} \BibitemShut
  {NoStop}%
\bibitem [{\citenamefont {Feinberg}\ and\ \citenamefont
  {Zee}(1997)}]{FEINBERG1997579}%
  \BibitemOpen
  \bibfield  {author} {\bibinfo {author} {\bibfnamefont {Joshua}\ \bibnamefont
  {Feinberg}}\ and\ \bibinfo {author} {\bibfnamefont {A.}~\bibnamefont {Zee}},\
  }\bibfield  {title} {\enquote {\bibinfo {title} {Non-hermitian random matrix
  theory: Method of hermitian reduction},}\ }\href {\doibase
  https://doi.org/10.1016/S0550-3213(97)00502-6} {\bibfield  {journal}
  {\bibinfo  {journal} {Nuclear Physics B}\ }\textbf {\bibinfo {volume}
  {504}},\ \bibinfo {pages} {579--608} (\bibinfo {year} {1997})}\BibitemShut
  {NoStop}%
\bibitem [{\citenamefont {Li}\ \emph {et~al.}(2017)\citenamefont {Li},
  \citenamefont {Liu}, \citenamefont {Xin},\ and\ \citenamefont
  {Zhou}}]{Li:2017hdt}%
  \BibitemOpen
  \bibfield  {author} {\bibinfo {author} {\bibfnamefont {Tianlin}\ \bibnamefont
  {Li}}, \bibinfo {author} {\bibfnamefont {Junyu}\ \bibnamefont {Liu}},
  \bibinfo {author} {\bibfnamefont {Yuan}\ \bibnamefont {Xin}}, \ and\ \bibinfo
  {author} {\bibfnamefont {Yehao}\ \bibnamefont {Zhou}},\ }\bibfield  {title}
  {\enquote {\bibinfo {title} {{Supersymmetric SYK model and random matrix
  theory}},}\ }\href {\doibase 10.1007/JHEP06(2017)111} {\bibfield  {journal}
  {\bibinfo  {journal} {JHEP}\ }\textbf {\bibinfo {volume} {06}},\ \bibinfo
  {pages} {111} (\bibinfo {year} {2017})}\BibitemShut {NoStop}%
\bibitem [{\citenamefont {S\'a}\ and\ \citenamefont
  {Garc\'\i{}a-Garc\'\i{}a}(2022)}]{Sa:2021rwg}%
  \BibitemOpen
  \bibfield  {author} {\bibinfo {author} {\bibfnamefont {Lucas}\ \bibnamefont
  {S\'a}}\ and\ \bibinfo {author} {\bibfnamefont {Antonio~M.}\ \bibnamefont
  {Garc\'\i{}a-Garc\'\i{}a}},\ }\bibfield  {title} {\enquote {\bibinfo {title}
  {{Q-Laguerre spectral density and quantum chaos in the
  Wishart-Sachdev-Ye-Kitaev model}},}\ }\href {\doibase
  10.1103/PhysRevD.105.026005} {\bibfield  {journal} {\bibinfo  {journal}
  {Phys. Rev. D}\ }\textbf {\bibinfo {volume} {105}},\ \bibinfo {pages}
  {026005} (\bibinfo {year} {2022})}\BibitemShut {NoStop}%
\bibitem [{\citenamefont {You}\ \emph {et~al.}(2017)\citenamefont {You},
  \citenamefont {Ludwig},\ and\ \citenamefont {Xu}}]{You:2016ldz}%
  \BibitemOpen
  \bibfield  {author} {\bibinfo {author} {\bibfnamefont {Yi-Zhuang}\
  \bibnamefont {You}}, \bibinfo {author} {\bibfnamefont {Andreas W.~W.}\
  \bibnamefont {Ludwig}}, \ and\ \bibinfo {author} {\bibfnamefont {Cenke}\
  \bibnamefont {Xu}},\ }\bibfield  {title} {\enquote {\bibinfo {title}
  {{Sachdev-Ye-Kitaev Model and Thermalization on the Boundary of Many-Body
  Localized Fermionic Symmetry Protected Topological States}},}\ }\href
  {\doibase 10.1103/PhysRevB.95.115150} {\bibfield  {journal} {\bibinfo
  {journal} {Phys. Rev. B}\ }\textbf {\bibinfo {volume} {95}},\ \bibinfo
  {pages} {115150} (\bibinfo {year} {2017})}\BibitemShut {NoStop}%
\bibitem [{\citenamefont {Gharibyan}\ \emph {et~al.}(2018)\citenamefont
  {Gharibyan}, \citenamefont {Hanada}, \citenamefont {Shenker},\ and\
  \citenamefont {Tezuka}}]{Gharibyan:2018jrp}%
  \BibitemOpen
  \bibfield  {author} {\bibinfo {author} {\bibfnamefont {Hrant}\ \bibnamefont
  {Gharibyan}}, \bibinfo {author} {\bibfnamefont {Masanori}\ \bibnamefont
  {Hanada}}, \bibinfo {author} {\bibfnamefont {Stephen~H.}\ \bibnamefont
  {Shenker}}, \ and\ \bibinfo {author} {\bibfnamefont {Masaki}\ \bibnamefont
  {Tezuka}},\ }\bibfield  {title} {\enquote {\bibinfo {title} {{Onset of Random
  Matrix Behavior in Scrambling Systems}},}\ }\href {\doibase
  10.1007/JHEP07(2018)124} {\bibfield  {journal} {\bibinfo  {journal} {JHEP}\
  }\textbf {\bibinfo {volume} {07}},\ \bibinfo {pages} {124} (\bibinfo {year}
  {2018})},\ \bibinfo {note} {[Erratum: JHEP 02, 197 (2019)]}\BibitemShut
  {NoStop}%
\bibitem [{\citenamefont {Nosaka}\ \emph {et~al.}(2018)\citenamefont {Nosaka},
  \citenamefont {Rosa},\ and\ \citenamefont {Yoon}}]{Nosaka:2018iat}%
  \BibitemOpen
  \bibfield  {author} {\bibinfo {author} {\bibfnamefont {Tomoki}\ \bibnamefont
  {Nosaka}}, \bibinfo {author} {\bibfnamefont {Dario}\ \bibnamefont {Rosa}}, \
  and\ \bibinfo {author} {\bibfnamefont {Junggi}\ \bibnamefont {Yoon}},\
  }\bibfield  {title} {\enquote {\bibinfo {title} {{The Thouless time for
  mass-deformed SYK}},}\ }\href {\doibase 10.1007/JHEP09(2018)041} {\bibfield
  {journal} {\bibinfo  {journal} {JHEP}\ }\textbf {\bibinfo {volume} {09}},\
  \bibinfo {pages} {041} (\bibinfo {year} {2018})}\BibitemShut {NoStop}%
\bibitem [{\citenamefont {Thouless}(1977)}]{PhysRevLett.39.1167}%
  \BibitemOpen
  \bibfield  {author} {\bibinfo {author} {\bibfnamefont {D.~J.}\ \bibnamefont
  {Thouless}},\ }\bibfield  {title} {\enquote {\bibinfo {title} {Maximum
  metallic resistance in thin wires},}\ }\href {\doibase
  10.1103/PhysRevLett.39.1167} {\bibfield  {journal} {\bibinfo  {journal}
  {Phys. Rev. Lett.}\ }\textbf {\bibinfo {volume} {39}},\ \bibinfo {pages}
  {1167--1169} (\bibinfo {year} {1977})}\BibitemShut {NoStop}%
\bibitem [{\citenamefont {Erd{\H{o}}s}\ and\ \citenamefont
  {Knowles}(2015)}]{Erdos2015}%
  \BibitemOpen
  \bibfield  {author} {\bibinfo {author} {\bibfnamefont {L{\'a}szl{\'o}}\
  \bibnamefont {Erd{\H{o}}s}}\ and\ \bibinfo {author} {\bibfnamefont {Antti}\
  \bibnamefont {Knowles}},\ }\bibfield  {title} {\enquote {\bibinfo {title}
  {{The Altshuler--Shklovskii Formulas for Random Band Matrices I: the
  Unimodular Case}},}\ }\href {\doibase 10.1007/s00220-014-2119-5} {\bibfield
  {journal} {\bibinfo  {journal} {Communications in Mathematical Physics}\
  }\textbf {\bibinfo {volume} {333}},\ \bibinfo {pages} {1365--1416} (\bibinfo
  {year} {2015})}\BibitemShut {NoStop}%
\bibitem [{\citenamefont {Prange}(1997)}]{PhysRevLett.78.2280}%
  \BibitemOpen
  \bibfield  {author} {\bibinfo {author} {\bibfnamefont {R.~E.}\ \bibnamefont
  {Prange}},\ }\bibfield  {title} {\enquote {\bibinfo {title} {The spectral
  form factor is not self-averaging},}\ }\href {\doibase
  10.1103/PhysRevLett.78.2280} {\bibfield  {journal} {\bibinfo  {journal}
  {Phys. Rev. Lett.}\ }\textbf {\bibinfo {volume} {78}},\ \bibinfo {pages}
  {2280--2283} (\bibinfo {year} {1997})}\BibitemShut {NoStop}%
\bibitem [{\citenamefont {Garc\'\i{}a-Garc\'\i{}a}\ \emph
  {et~al.}(2018{\natexlab{a}})\citenamefont {Garc\'\i{}a-Garc\'\i{}a},
  \citenamefont {Jia},\ and\ \citenamefont
  {Verbaarschot}}]{Garcia-Garcia:2018ruf}%
  \BibitemOpen
  \bibfield  {author} {\bibinfo {author} {\bibfnamefont {Antonio~M.}\
  \bibnamefont {Garc\'\i{}a-Garc\'\i{}a}}, \bibinfo {author} {\bibfnamefont
  {Yiyang}\ \bibnamefont {Jia}}, \ and\ \bibinfo {author} {\bibfnamefont
  {Jacobus J.~M.}\ \bibnamefont {Verbaarschot}},\ }\bibfield  {title} {\enquote
  {\bibinfo {title} {{Universality and Thouless energy in the supersymmetric
  Sachdev-Ye-Kitaev Model}},}\ }\href {\doibase 10.1103/PhysRevD.97.106003}
  {\bibfield  {journal} {\bibinfo  {journal} {Phys. Rev. D}\ }\textbf {\bibinfo
  {volume} {97}},\ \bibinfo {pages} {106003} (\bibinfo {year}
  {2018}{\natexlab{a}})}\BibitemShut {NoStop}%
\bibitem [{\citenamefont {Matsoukas-Roubeas}\ \emph
  {et~al.}(2023{\natexlab{b}})\citenamefont {Matsoukas-Roubeas}, \citenamefont
  {Beau}, \citenamefont {Santos},\ and\ \citenamefont {del
  Campo}}]{Matsoukas-Roubeas:2023xge}%
  \BibitemOpen
  \bibfield  {author} {\bibinfo {author} {\bibfnamefont {Apollonas~S.}\
  \bibnamefont {Matsoukas-Roubeas}}, \bibinfo {author} {\bibfnamefont
  {Mathieu}\ \bibnamefont {Beau}}, \bibinfo {author} {\bibfnamefont {Lea~F.}\
  \bibnamefont {Santos}}, \ and\ \bibinfo {author} {\bibfnamefont {Adolfo}\
  \bibnamefont {del Campo}},\ }\bibfield  {title} {\enquote {\bibinfo {title}
  {{Unitarity breaking in self-averaging spectral form factors}},}\ }\href
  {\doibase 10.1103/PhysRevA.108.062201} {\bibfield  {journal} {\bibinfo
  {journal} {Phys. Rev. A}\ }\textbf {\bibinfo {volume} {108}},\ \bibinfo
  {pages} {062201} (\bibinfo {year} {2023}{\natexlab{b}})}\BibitemShut
  {NoStop}%
\bibitem [{sup()}]{supp}%
  \BibitemOpen
  \href@noop {} {}\bibinfo {note} {See Supplemental Materials with additional
  references \cite{Giraud:2020mmb,Iyoda:2018osm} at
  URL-will-be-inserted-by-publisher for detailed information on the comparison
  between singular values and eigenvalues and the behavior of the singular
  complexity.}\BibitemShut {Stop}%
\bibitem [{\citenamefont {Maldacena}(1998)}]{Maldacena:1997re}%
  \BibitemOpen
  \bibfield  {author} {\bibinfo {author} {\bibfnamefont {Juan~Martin}\
  \bibnamefont {Maldacena}},\ }\bibfield  {title} {\enquote {\bibinfo {title}
  {{The Large N limit of superconformal field theories and supergravity}},}\
  }\href {\doibase 10.4310/ATMP.1998.v2.n2.a1} {\bibfield  {journal} {\bibinfo
  {journal} {Adv. Theor. Math. Phys.}\ }\textbf {\bibinfo {volume} {2}},\
  \bibinfo {pages} {231--252} (\bibinfo {year} {1998})}\BibitemShut {NoStop}%
\bibitem [{\citenamefont {Witten}(1998)}]{Witten:1998qj}%
  \BibitemOpen
  \bibfield  {author} {\bibinfo {author} {\bibfnamefont {Edward}\ \bibnamefont
  {Witten}},\ }\bibfield  {title} {\enquote {\bibinfo {title} {{Anti-de Sitter
  space and holography}},}\ }\href {\doibase 10.4310/ATMP.1998.v2.n2.a2}
  {\bibfield  {journal} {\bibinfo  {journal} {Adv. Theor. Math. Phys.}\
  }\textbf {\bibinfo {volume} {2}},\ \bibinfo {pages} {253--291} (\bibinfo
  {year} {1998})}\BibitemShut {NoStop}%
\bibitem [{\citenamefont {Jefferson}\ and\ \citenamefont
  {Myers}(2017)}]{Jefferson:2017sdb}%
  \BibitemOpen
  \bibfield  {author} {\bibinfo {author} {\bibfnamefont {Ro}~\bibnamefont
  {Jefferson}}\ and\ \bibinfo {author} {\bibfnamefont {Robert~C.}\ \bibnamefont
  {Myers}},\ }\bibfield  {title} {\enquote {\bibinfo {title} {{Circuit
  complexity in quantum field theory}},}\ }\href {\doibase
  10.1007/JHEP10(2017)107} {\bibfield  {journal} {\bibinfo  {journal} {JHEP}\
  }\textbf {\bibinfo {volume} {10}},\ \bibinfo {pages} {107} (\bibinfo {year}
  {2017})}\BibitemShut {NoStop}%
\bibitem [{\citenamefont {Balasubramanian}\ \emph {et~al.}(2022)\citenamefont
  {Balasubramanian}, \citenamefont {Caputa}, \citenamefont {Magan},\ and\
  \citenamefont {Wu}}]{Balasubramanian:2022tpr}%
  \BibitemOpen
  \bibfield  {author} {\bibinfo {author} {\bibfnamefont {Vijay}\ \bibnamefont
  {Balasubramanian}}, \bibinfo {author} {\bibfnamefont {Pawel}\ \bibnamefont
  {Caputa}}, \bibinfo {author} {\bibfnamefont {Javier~M.}\ \bibnamefont
  {Magan}}, \ and\ \bibinfo {author} {\bibfnamefont {Qingyue}\ \bibnamefont
  {Wu}},\ }\bibfield  {title} {\enquote {\bibinfo {title} {{Quantum chaos and
  the complexity of spread of states}},}\ }\href {\doibase
  10.1103/PhysRevD.106.046007} {\bibfield  {journal} {\bibinfo  {journal}
  {Phys. Rev. D}\ }\textbf {\bibinfo {volume} {106}},\ \bibinfo {pages}
  {046007} (\bibinfo {year} {2022})}\BibitemShut {NoStop}%
\bibitem [{\citenamefont {Erdmenger}\ \emph {et~al.}(2023)\citenamefont
  {Erdmenger}, \citenamefont {Jian},\ and\ \citenamefont
  {Xian}}]{Erdmenger:2023wjg}%
  \BibitemOpen
  \bibfield  {author} {\bibinfo {author} {\bibfnamefont {Johanna}\ \bibnamefont
  {Erdmenger}}, \bibinfo {author} {\bibfnamefont {Shao-Kai}\ \bibnamefont
  {Jian}}, \ and\ \bibinfo {author} {\bibfnamefont {Zhuo-Yu}\ \bibnamefont
  {Xian}},\ }\bibfield  {title} {\enquote {\bibinfo {title} {{Universal chaotic
  dynamics from Krylov space}},}\ }\href {\doibase 10.1007/JHEP08(2023)176}
  {\bibfield  {journal} {\bibinfo  {journal} {JHEP}\ }\textbf {\bibinfo
  {volume} {08}},\ \bibinfo {pages} {176} (\bibinfo {year} {2023})}\BibitemShut
  {NoStop}%
\bibitem [{\citenamefont {Parker}\ \emph {et~al.}(2019)\citenamefont {Parker},
  \citenamefont {Cao}, \citenamefont {Avdoshkin}, \citenamefont {Scaffidi},\
  and\ \citenamefont {Altman}}]{Parker:2018yvk}%
  \BibitemOpen
  \bibfield  {author} {\bibinfo {author} {\bibfnamefont {Daniel~E.}\
  \bibnamefont {Parker}}, \bibinfo {author} {\bibfnamefont {Xiangyu}\
  \bibnamefont {Cao}}, \bibinfo {author} {\bibfnamefont {Alexander}\
  \bibnamefont {Avdoshkin}}, \bibinfo {author} {\bibfnamefont {Thomas}\
  \bibnamefont {Scaffidi}}, \ and\ \bibinfo {author} {\bibfnamefont {Ehud}\
  \bibnamefont {Altman}},\ }\bibfield  {title} {\enquote {\bibinfo {title} {{A
  Universal Operator Growth Hypothesis}},}\ }\href {\doibase
  10.1103/PhysRevX.9.041017} {\bibfield  {journal} {\bibinfo  {journal} {Phys.
  Rev. X}\ }\textbf {\bibinfo {volume} {9}},\ \bibinfo {pages} {041017}
  (\bibinfo {year} {2019})}\BibitemShut {NoStop}%
\bibitem [{\citenamefont {Camargo}\ \emph
  {et~al.}(2024{\natexlab{a}})\citenamefont {Camargo}, \citenamefont {Jahnke},
  \citenamefont {Jeong}, \citenamefont {Kim},\ and\ \citenamefont
  {Nishida}}]{Camargo:2023eev}%
  \BibitemOpen
  \bibfield  {author} {\bibinfo {author} {\bibfnamefont {Hugo~A.}\ \bibnamefont
  {Camargo}}, \bibinfo {author} {\bibfnamefont {Viktor}\ \bibnamefont
  {Jahnke}}, \bibinfo {author} {\bibfnamefont {Hyun-Sik}\ \bibnamefont
  {Jeong}}, \bibinfo {author} {\bibfnamefont {Keun-Young}\ \bibnamefont {Kim}},
  \ and\ \bibinfo {author} {\bibfnamefont {Mitsuhiro}\ \bibnamefont
  {Nishida}},\ }\bibfield  {title} {\enquote {\bibinfo {title} {{Spectral and
  Krylov complexity in billiard systems}},}\ }\href {\doibase
  10.1103/PhysRevD.109.046017} {\bibfield  {journal} {\bibinfo  {journal}
  {Phys. Rev. D}\ }\textbf {\bibinfo {volume} {109}},\ \bibinfo {pages}
  {046017} (\bibinfo {year} {2024}{\natexlab{a}})}\BibitemShut {NoStop}%
\bibitem [{\citenamefont {Camargo}\ \emph
  {et~al.}(2024{\natexlab{b}})\citenamefont {Camargo}, \citenamefont {Huh},
  \citenamefont {Jahnke}, \citenamefont {Jeong}, \citenamefont {Kim},\ and\
  \citenamefont {Nishida}}]{Camargo:2024deu}%
  \BibitemOpen
  \bibfield  {author} {\bibinfo {author} {\bibfnamefont {Hugo~A.}\ \bibnamefont
  {Camargo}}, \bibinfo {author} {\bibfnamefont {Kyoung-Bum}\ \bibnamefont
  {Huh}}, \bibinfo {author} {\bibfnamefont {Viktor}\ \bibnamefont {Jahnke}},
  \bibinfo {author} {\bibfnamefont {Hyun-Sik}\ \bibnamefont {Jeong}}, \bibinfo
  {author} {\bibfnamefont {Keun-Young}\ \bibnamefont {Kim}}, \ and\ \bibinfo
  {author} {\bibfnamefont {Mitsuhiro}\ \bibnamefont {Nishida}},\ }\bibfield
  {title} {\enquote {\bibinfo {title} {{Spread and spectral complexity in
  quantum spin chains: from integrability to chaos}},}\ }\href {\doibase
  10.1007/JHEP08(2024)241} {\bibfield  {journal} {\bibinfo  {journal} {JHEP}\
  }\textbf {\bibinfo {volume} {08}},\ \bibinfo {pages} {241} (\bibinfo {year}
  {2024}{\natexlab{b}})}\BibitemShut {NoStop}%
\bibitem [{\citenamefont {Garc\'\i{}a-Garc\'\i{}a}\ \emph
  {et~al.}(2023{\natexlab{b}})\citenamefont {Garc\'\i{}a-Garc\'\i{}a},
  \citenamefont {S\'a}, \citenamefont {Verbaarschot},\ and\ \citenamefont
  {Zheng}}]{Garcia-Garcia:2022adg}%
  \BibitemOpen
  \bibfield  {author} {\bibinfo {author} {\bibfnamefont {Antonio~M.}\
  \bibnamefont {Garc\'\i{}a-Garc\'\i{}a}}, \bibinfo {author} {\bibfnamefont
  {Lucas}\ \bibnamefont {S\'a}}, \bibinfo {author} {\bibfnamefont {Jacobus
  J.~M.}\ \bibnamefont {Verbaarschot}}, \ and\ \bibinfo {author} {\bibfnamefont
  {Jie~Ping}\ \bibnamefont {Zheng}},\ }\bibfield  {title} {\enquote {\bibinfo
  {title} {{Keldysh wormholes and anomalous relaxation in the dissipative
  Sachdev-Ye-Kitaev model}},}\ }\href {\doibase 10.1103/PhysRevD.107.106006}
  {\bibfield  {journal} {\bibinfo  {journal} {Phys. Rev. D}\ }\textbf {\bibinfo
  {volume} {107}},\ \bibinfo {pages} {106006} (\bibinfo {year}
  {2023}{\natexlab{b}})}\BibitemShut {NoStop}%
\bibitem [{\citenamefont {Cai}\ \emph {et~al.}(2022)\citenamefont {Cai},
  \citenamefont {Cao}, \citenamefont {Ge}, \citenamefont {Matsumoto},\ and\
  \citenamefont {Sin}}]{Cai:2022onu}%
  \BibitemOpen
  \bibfield  {author} {\bibinfo {author} {\bibfnamefont {Wenhe}\ \bibnamefont
  {Cai}}, \bibinfo {author} {\bibfnamefont {Sizheng}\ \bibnamefont {Cao}},
  \bibinfo {author} {\bibfnamefont {Xian-Hui}\ \bibnamefont {Ge}}, \bibinfo
  {author} {\bibfnamefont {Masataka}\ \bibnamefont {Matsumoto}}, \ and\
  \bibinfo {author} {\bibfnamefont {Sang-Jin}\ \bibnamefont {Sin}},\ }\bibfield
   {title} {\enquote {\bibinfo {title} {{Non-Hermitian quantum system generated
  from two coupled Sachdev-Ye-Kitaev models}},}\ }\href {\doibase
  10.1103/PhysRevD.106.106010} {\bibfield  {journal} {\bibinfo  {journal}
  {Phys. Rev. D}\ }\textbf {\bibinfo {volume} {106}},\ \bibinfo {pages}
  {106010} (\bibinfo {year} {2022})}\BibitemShut {NoStop}%
\bibitem [{\citenamefont {Cao}\ and\ \citenamefont {Ge}(2024)}]{Cao:2024xaw}%
  \BibitemOpen
  \bibfield  {author} {\bibinfo {author} {\bibfnamefont {Sizheng}\ \bibnamefont
  {Cao}}\ and\ \bibinfo {author} {\bibfnamefont {Xian-Hui}\ \bibnamefont
  {Ge}},\ }\bibfield  {title} {\enquote {\bibinfo {title} {{Excitation
  transmission through a non-Hermitian traversable wormhole}},}\ }\href
  {\doibase 10.1103/PhysRevD.110.046022} {\bibfield  {journal} {\bibinfo
  {journal} {Phys. Rev. D}\ }\textbf {\bibinfo {volume} {110}},\ \bibinfo
  {pages} {046022} (\bibinfo {year} {2024})}\BibitemShut {NoStop}%
\bibitem [{\citenamefont {Garc\'\i{}a-Garc\'\i{}a}\ \emph
  {et~al.}(2018{\natexlab{b}})\citenamefont {Garc\'\i{}a-Garc\'\i{}a},
  \citenamefont {Loureiro}, \citenamefont {Romero-Berm\'udez},\ and\
  \citenamefont {Tezuka}}]{Garcia-Garcia:2017bkg}%
  \BibitemOpen
  \bibfield  {author} {\bibinfo {author} {\bibfnamefont {Antonio~M.}\
  \bibnamefont {Garc\'\i{}a-Garc\'\i{}a}}, \bibinfo {author} {\bibfnamefont
  {Bruno}\ \bibnamefont {Loureiro}}, \bibinfo {author} {\bibfnamefont
  {Aurelio}\ \bibnamefont {Romero-Berm\'udez}}, \ and\ \bibinfo {author}
  {\bibfnamefont {Masaki}\ \bibnamefont {Tezuka}},\ }\bibfield  {title}
  {\enquote {\bibinfo {title} {{Chaotic-Integrable Transition in the
  Sachdev-Ye-Kitaev Model}},}\ }\href {\doibase 10.1103/PhysRevLett.120.241603}
  {\bibfield  {journal} {\bibinfo  {journal} {Phys. Rev. Lett.}\ }\textbf
  {\bibinfo {volume} {120}},\ \bibinfo {pages} {241603} (\bibinfo {year}
  {2018}{\natexlab{b}})}\BibitemShut {NoStop}%
\bibitem [{\citenamefont {Srivastava}\ \emph {et~al.}(2018)\citenamefont
  {Srivastava}, \citenamefont {Lakshminarayan}, \citenamefont {Tomsovic},\ and\
  \citenamefont {Bäcker}}]{Srivastava_2019}%
  \BibitemOpen
  \bibfield  {author} {\bibinfo {author} {\bibfnamefont {Shashi C~L}\
  \bibnamefont {Srivastava}}, \bibinfo {author} {\bibfnamefont {Arul}\
  \bibnamefont {Lakshminarayan}}, \bibinfo {author} {\bibfnamefont {Steven}\
  \bibnamefont {Tomsovic}}, \ and\ \bibinfo {author} {\bibfnamefont {Arnd}\
  \bibnamefont {Bäcker}},\ }\bibfield  {title} {\enquote {\bibinfo {title}
  {Ordered level spacing probability densities},}\ }\href {\doibase
  10.1088/1751-8121/aaefa4} {\bibfield  {journal} {\bibinfo  {journal} {Journal
  of Physics A: Mathematical and Theoretical}\ }\textbf {\bibinfo {volume}
  {52}},\ \bibinfo {pages} {025101} (\bibinfo {year} {2018})}\BibitemShut
  {NoStop}%
\bibitem [{\citenamefont {Tekur}\ \emph {et~al.}(2018)\citenamefont {Tekur},
  \citenamefont {Bhosale},\ and\ \citenamefont
  {Santhanam}}]{Santhanamhigherspacing18}%
  \BibitemOpen
  \bibfield  {author} {\bibinfo {author} {\bibfnamefont {S.~Harshini}\
  \bibnamefont {Tekur}}, \bibinfo {author} {\bibfnamefont {Udaysinh~T.}\
  \bibnamefont {Bhosale}}, \ and\ \bibinfo {author} {\bibfnamefont {M.~S.}\
  \bibnamefont {Santhanam}},\ }\bibfield  {title} {\enquote {\bibinfo {title}
  {{Higher-order spacing ratios in random matrix theory and complex quantum
  systems}},}\ }\href {\doibase 10.1103/physrevb.98.104305} {\bibfield
  {journal} {\bibinfo  {journal} {Phys. Rev. B}\ }\textbf {\bibinfo {volume}
  {98}},\ \bibinfo {pages} {104305} (\bibinfo {year} {2018})}\BibitemShut
  {NoStop}%
\bibitem [{\citenamefont {Shir}\ \emph {et~al.}(2023)\citenamefont {Shir},
  \citenamefont {Martinez-Azcona},\ and\ \citenamefont {Chenu}}]{Shir:2023olc}%
  \BibitemOpen
  \bibfield  {author} {\bibinfo {author} {\bibfnamefont {Ruth}\ \bibnamefont
  {Shir}}, \bibinfo {author} {\bibfnamefont {Pablo}\ \bibnamefont
  {Martinez-Azcona}}, \ and\ \bibinfo {author} {\bibfnamefont {Aur\'elia}\
  \bibnamefont {Chenu}},\ }\bibfield  {title} {\enquote {\bibinfo {title}
  {{Full range spectral correlations and their spectral form factors in chaotic
  and integrable models}},}\ }\href@noop {} {\  (\bibinfo {year} {2023})},\
  \Eprint {http://arxiv.org/abs/2311.09292} {arXiv:2311.09292 [quant-ph]}
  \BibitemShut {NoStop}%
\bibitem [{\citenamefont {Evers}\ and\ \citenamefont
  {Mirlin}(2008)}]{RevModPhys.80.1355}%
  \BibitemOpen
  \bibfield  {author} {\bibinfo {author} {\bibfnamefont {Ferdinand}\
  \bibnamefont {Evers}}\ and\ \bibinfo {author} {\bibfnamefont {Alexander~D.}\
  \bibnamefont {Mirlin}},\ }\bibfield  {title} {\enquote {\bibinfo {title}
  {Anderson transitions},}\ }\href {\doibase 10.1103/RevModPhys.80.1355}
  {\bibfield  {journal} {\bibinfo  {journal} {Rev. Mod. Phys.}\ }\textbf
  {\bibinfo {volume} {80}},\ \bibinfo {pages} {1355--1417} (\bibinfo {year}
  {2008})}\BibitemShut {NoStop}%
\bibitem [{\citenamefont {Garc\'{\i}a-Garc\'{\i}a}\ and\ \citenamefont
  {Verbaarschot}(2016)}]{PhysRevD.94.126010}%
  \BibitemOpen
  \bibfield  {author} {\bibinfo {author} {\bibfnamefont {Antonio~M.}\
  \bibnamefont {Garc\'{\i}a-Garc\'{\i}a}}\ and\ \bibinfo {author}
  {\bibfnamefont {Jacobus J.~M.}\ \bibnamefont {Verbaarschot}},\ }\bibfield
  {title} {\enquote {\bibinfo {title} {Spectral and thermodynamic properties of
  the {Sachdev-Ye-Kitaev} model},}\ }\href {\doibase
  10.1103/PhysRevD.94.126010} {\bibfield  {journal} {\bibinfo  {journal} {Phys.
  Rev. D}\ }\textbf {\bibinfo {volume} {94}},\ \bibinfo {pages} {126010}
  (\bibinfo {year} {2016})}\BibitemShut {NoStop}%
\bibitem [{\citenamefont {Giraud}\ \emph {et~al.}(2022)\citenamefont {Giraud},
  \citenamefont {Mac\'e}, \citenamefont {Vernier},\ and\ \citenamefont
  {Alet}}]{Giraud:2020mmb}%
  \BibitemOpen
  \bibfield  {author} {\bibinfo {author} {\bibfnamefont {Olivier}\ \bibnamefont
  {Giraud}}, \bibinfo {author} {\bibfnamefont {Nicolas}\ \bibnamefont
  {Mac\'e}}, \bibinfo {author} {\bibfnamefont {\'Eric}\ \bibnamefont
  {Vernier}}, \ and\ \bibinfo {author} {\bibfnamefont {Fabien}\ \bibnamefont
  {Alet}},\ }\bibfield  {title} {\enquote {\bibinfo {title} {{Probing
  Symmetries of Quantum Many-Body Systems through Gap Ratio Statistics}},}\
  }\href {\doibase 10.1103/PhysRevX.12.011006} {\bibfield  {journal} {\bibinfo
  {journal} {Phys. Rev. X}\ }\textbf {\bibinfo {volume} {12}},\ \bibinfo
  {pages} {011006} (\bibinfo {year} {2022})}\BibitemShut {NoStop}%
\bibitem [{\citenamefont {Iyoda}\ \emph {et~al.}(2018)\citenamefont {Iyoda},
  \citenamefont {Katsura},\ and\ \citenamefont {Sagawa}}]{Iyoda:2018osm}%
  \BibitemOpen
  \bibfield  {author} {\bibinfo {author} {\bibfnamefont {Eiki}\ \bibnamefont
  {Iyoda}}, \bibinfo {author} {\bibfnamefont {Hosho}\ \bibnamefont {Katsura}},
  \ and\ \bibinfo {author} {\bibfnamefont {Takahiro}\ \bibnamefont {Sagawa}},\
  }\bibfield  {title} {\enquote {\bibinfo {title} {{Effective dimension, level
  statistics, and integrability of Sachdev-Ye-Kitaev-like models}},}\ }\href
  {\doibase 10.1103/PhysRevD.98.086020} {\bibfield  {journal} {\bibinfo
  {journal} {Phys. Rev. D}\ }\textbf {\bibinfo {volume} {98}},\ \bibinfo
  {pages} {086020} (\bibinfo {year} {2018})}\BibitemShut {NoStop}%
\end{thebibliography}%

\newpage

\hbox{}\thispagestyle{empty}\newpage
\appendix\label{appendix}
\onecolumngrid
\begin{center}
\textbf{\large Supplemental Materials: \\Probing quantum chaos through singular-value correlations \\in sparse non-Hermitian SYK model}
\vspace{2ex}

Pratik Nandy, Tanay Pathak, and Masaki Tezuka
\end{center}

\section{S1. From $\upsigma$FF to SFF: Reduction to Hermitian systems}\label{sec:Test}

In Hermitian systems, a key observation is that the singular values correspond to the absolute values of the eigenvalues. This distinction is crucial, as it implies that the SFF and $\upsigma$FF, exhibit fundamentally different behaviors. This difference is illustrated in the inset of Fig.\,\ref{fig:Thoulesstimevsp}, which depicts the dynamics of SFF and $\upsigma$FF for the dense Hermitian SYK$_4$ model. The model is derived by setting $x_{abcd}=1$ and $M_{abcd} = 0$ for all $\{a,b,c,d\}$  in its non-Hermitian counterpart (Eq.\,\eqref{nhsykhsparse}). Replacing $\sigma_n$ by eigenvalues $E_n$ in Eq.\,\eqref{gaussfilter0}, we get the analogous definition of filtered SFF.

\begin{table}[h]
\hspace*{-0.1 cm}
\begin{tabular}{|l|l|l|l|l|l|l|l|}
\hline
~~~~~~$\mathrm{System}$   & ~~$\braket{r_\upsigma}$ & ~~$\braket{r}$& \,~$\braket{r_{+}}$  & \,~$\braket{r_{-}}$ & $\braket{r}$ (2 blocks) \\ \hline
nSYK ($N = 20$)& \textcolor{blue} {0.4119}   &  0.6743  &  0.6744  &  0.6744 & \textcolor{blue}{0.4117} \\ \hline
nSYK ($N = 22$) & \textcolor{blue}{0.4220}   & 0.5994   & 0.5998   & 0.5991 & \textcolor{blue}{0.4220}   \\ \hline
nSYK ($N = 24$) &   \textcolor{blue}{0.4238}  & 0.5303   & 0.5303   & 0.5302  & \textcolor{blue}{0.4234}   \\ \hline
\end{tabular}
\caption{The Table shows the comparison between the singular value $\braket{r_\upsigma}$-value and the $\braket{r}$-values for the \emph{dense} Hermitian SYK model for different system sizes $N = 20$ (GSE), $N = 22$ (GUE), $N = 24$ (GOE). We take 1000 Hamiltonian realizations for each case. The Hermitian model is obtained by setting $x_{abcd}=1$ and $M_{abcd} = 0$ for all $\{a,b,c,d\}$ in the Hamiltonian \eqref{nhsykhsparse}. Here the $\braket{r_{+}}$-value includes the positive eigenvalues while $\braket{r_{-}}$-value includes the negative eigenvalues only. They are to be matched with the $\braket{r}_{\mathrm{RMT}}$ values in Table \ref{Tab1} and \cite{Atas2013distribution}. The $\braket{r_\upsigma}$-values are compared with the $\braket{r}$-value of $2$ blocks (marked in \textcolor{blue}{blue}) \cite{Orman:2024mpw}. Explanation is given in the text.} \label{Tab2}
\end{table}

Notably, the singular values offer a \emph{weaker} effect compared to the eigenvalues, considering the ramp. However, to reduce the SFF from $\upsigma$FF, we consider the following prescription. From SVD, we obtain two sets of vectors, the left and right singular vectors. They are related by
\begin{align}
    H \ket{u_n} = \sigma_n \ket{v_n}\,, ~~~~ H^{\dagger} \ket{v_n} = \sigma_n \ket{u_n}\,,
\end{align}
where $\{\sigma_n\}$ are the singular values. For Hermitian systems, the eigenvalues are real, and the left ($\ket{u_n}$) and the right ($\ket{v_n}$) singular vectors are related with a factor of $\pm 1$. If the left and right singular vectors are equal, i.e., $\ket{u_n} = \ket{v_n}$ for some $n$, then the corresponding singular value exactly equals the eigenvalue i.e., $\sigma_n = E_n$.  However, if they differ by a negative sign i.e., $\ket{u_n} = - \ket{v_n}$ for some $n$, then the corresponding singular value and the eigenvalue differs by a negative sign i.e., $\sigma_n = - E_n$. It is thus straightforward to identify such vectors such that the singular values can be directly mapped with the eigenvalues, and correspondingly compute the $\upsigma$FF. In this case, the $\upsigma$FF exactly equals the SFF as shown by the red line in the inset of Fig.\,\ref{fig:Thoulesstimevsp}.

Intriguingly, the Hermitian system exhibits a unique characteristic in its $\braket{r_\upsigma}$-value. As depicted in Table \ref{Tab2}, there is a notable distinction when comparing the $\braket{r_\upsigma}$-value to the $\braket{r}$-values within the identical Hermitian SYK$_4$ model. Additionally, the $\braket{r_{+}}$-value and $\braket{r_{-}}$-value were computed separately, considering only the positive and negative eigenvalues of the Hamiltonian, respectively. These ratios align with the overall $\braket{r}$-value, which is anticipated since it solely accounts for the spacing between consecutive eigenvalues. Nonetheless, these individual ratios as well as the overall $\braket{r}$-value do not correspond to the $\braket{r_\upsigma}$-value, which is identified as the $\braket{r}$-value for two separate blocks before symmetry resolution. This discrepancy arises because, in Hermitian systems, the singular values are determined by the absolute magnitudes of the eigenvalues, leading to the computation of the $\braket{r_\upsigma}$-value mirroring the $\braket{r}$-value for bifurcated blocks \cite{Giraud:2020mmb, KawabataSVD23, Orman:2024mpw}. We anticipate such bifurcation has a similar effect of having the presence of conserved charges \cite{Iyoda:2018osm, Sa:2021rwg}.  Curiously, this phenomenon is exclusive to Hermitian and anti-Hermitian (with $J_{abcd} = 0$, $M_{abcd} \neq 0$) systems and echoes the discrepancy between the $\upsigma$FF and SFF as shown in the inset of Fig.\,\ref{fig:Thoulesstimevsp}.

\section{S2. Behavior of singular complexity}\label{sec:scapp}

\begin{figure}[t]
\hspace*{-0.9 cm}
\includegraphics[width=0.41\textwidth]{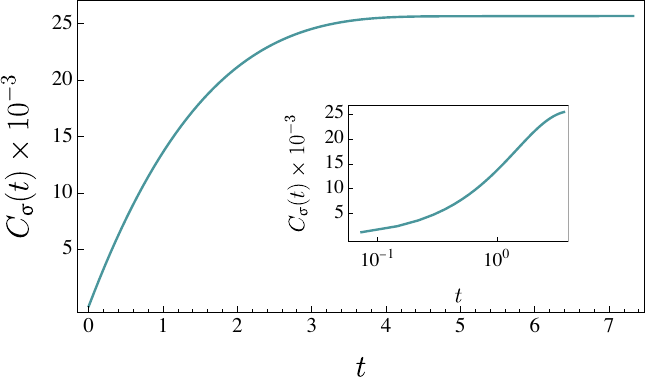}
\caption{Behaviour of the singular complexity with time for \emph{dense} ($p=1$) non-Hermitian SYK model. The system parameter is $N =26$ with $1000$ random Hamiltonian realizations taken. Inset shows the early time behavior of the complexity, which increases quadratically followed by linear growth and saturation.} \label{fig:SpecCN26wtime}
\end{figure}

Here we present the detailed behavior of singular complexity \cite{Iliesiu:2021ari}. It is defined at finite temperature $T = 1/\beta$ as 
\begin{align}
        C_{\upsigma}(t) = \frac{1}{Z_{2\upsigma}(\beta) L} \sum_{\sigma_i \neq \sigma_j} \left[\frac{\sin (t (\sigma_i - \sigma_j)/2)}{(\sigma_i - \sigma_j)/2}\right]^2 e^{-\beta (\sigma_i + \sigma_j)}\,. \label{singC}
\end{align}
Here $Z_{\upsigma}(\beta) = \sum_n e^{-\beta \sigma_n}$ is the thermal \emph{singular partition function}, constructed from the singular values of the corresponding non-Hermitian Hamiltonian. In the case of degeneracies, we keep the degenerate spectrum. The derivative of the singular complexity is related to the $\upsigma$FF as
\begin{align}
    \frac{d^2}{dt^2}  C_{\upsigma}(t) = \frac{2}{L} \frac{Z_{\upsigma}(\beta)^2}{Z_{\upsigma}(2 \beta)} \upsigma \mathrm{FF}(t) - \frac{2}{L}\,,
\end{align}
in a similar spirit to spectral complexity \cite{Iliesiu:2021ari, Erdmenger:2023wjg} which can be checked. Focusing on the infinite temperature $\beta = 0$ limit, \eqref{singC} takes a simpler form as in the main text:
\begin{align}
    C_{\upsigma}(t) = \frac{1}{L^2} \sum_{\sigma_i \neq \sigma_j} \left[\frac{\sin (t (\sigma_i - \sigma_j)/2)}{(\sigma_i - \sigma_j)/2}\right]^2\,.  \tag{\ref{scTinf}}
\end{align}
At early times, it grows quadratically as $C_{\upsigma}(t) \approx (1-1/L) t^2$, followed by a linear growth and plateau regime. Figure \ref{fig:SpecCN26wtime} shows the time evolution of the infinite-temperature singular complexity \eqref{scTinf} for the \emph{dense} Hamiltonian for $N = 26$ and $1000$ realizations. The complexity grows quadratically, followed by linear growth (inset) and saturation. The saturation value is elevated as the sparsity increases. 

\end{document}